\documentclass{aa}
\usepackage{graphicx}
\usepackage[varg]{txfonts}
\usepackage{natbib,twoopt}
\usepackage[breaklinks=true]{hyperref} 
\bibpunct{(}{)}{;}{a}{}{,} 
\makeatletter
\newcommandtwoopt{\citeads}[3][][]{\href{http://adsabs.harvard.edu/abs/#3}%
{\def\hyper@linkstart##1##2{}%
\let\hyper@linkend\@empty\citealp[#1][#2]{#3}}}
\newcommandtwoopt{\citepads}[3][][]{\href{http://adsabs.harvard.edu/abs/#3}%
{\def\hyper@linkstart##1##2{}%
\let\hyper@linkend\@empty\citep[#1][#2]{#3}}}
\newcommandtwoopt{\citetads}[3][][]{\href{http://adsabs.harvard.edu/abs/#3}%
{\def\hyper@linkstart##1##2{}%
\let\hyper@linkend\@empty\citet[#1][#2]{#3}}}
\newcommandtwoopt{\citeyearads}[3][][]%
{\href{http://adsabs.harvard.edu/abs/#3}
{\def\hyper@linkstart##1##2{}%
\let\hyper@linkend\@empty\citeyear[#1][#2]{#3}}}
\makeatother

\usepackage{lscape}
\usepackage{xcolor}

\begin{document} 
   \title{Convective blueshift strengths for 242 evolved stars\thanks{Table \ref{tab:cheatsheet} and the full Table \ref{tab:star_data} are available in electronic form at the CDS via anonymous ftp to \url{cdsarc.u-strasbg.fr} (130.79.128.5) or via \url{http://cdsweb.u-strasbg.fr/cgi-bin/qcat?J/A+A/}}}

   \author{F. Liebing\inst{\ref{inst1}, \ref{inst2}} \and S. V. Jeffers\inst{\ref{inst1}} \and M. Zechmeister\inst{\ref{inst2}} \and A. Reiners\inst{\ref{inst2}} }

   \institute{Max Planck Institute for Solar System Research,
              Justus-von-Liebig-Weg 3, 37077 G\"ottingen, Germany\\
              \email{liebing@mps.mpg.de}\label{inst1}
              \and
              Institut f\"ur Astrophysik und Geophysik (IAG), Universit\"at G\"ottingen,
              Friedrich-Hund-Platz 1, 37077 G\"ottingen, Germany\\
              \label{inst2}
	     }

   \date{Received xxx /	Accepted xxx}
 
  \abstract
   {With the advent of extreme precision radial velocity (RV) surveys, seeking to detect planets at RV semi-amplitudes of 10\,cm\,s$^{-1}$, intrinsic stellar variability is the biggest challenge to detecting small exoplanets. To overcome the challenge we must first thoroughly understand all facets of stellar variability. Among those, convective blueshift caused by stellar granulation and its suppression through magnetic activity plays a significant role in covering planetary signals in stellar jitter.}
   {Previously we found that for main sequence stars, convective blueshift as an observational proxy for the strength of convection near the stellar surface strongly depends on effective temperature. In this work we investigate 242 post main sequence stars, covering the subgiant, red giant, and asymptotic giant phases and empirically determine the changes in convective blueshift with advancing stellar evolution.}
   {We used the third signature scaling approach to fit a solar model for the convective blueshift to absorption-line shift measurements from a sample of coadded HARPS spectra, ranging in temperature from 3750\,K to 6150\,K. We compare the results to main sequence stars of comparable temperatures but with a higher surface gravity.}
   {We show that convective blueshift becomes significantly stronger for evolved stars compared to main sequence stars of a similar temperature. The difference increases as the star becomes more evolved, reaching a 5x increase below 4300\,K for the most evolved stars. The large number of stars in the sample, for the first time, allowed for us to empirically show that convective blueshift remains almost constant among the entire evolved star sample at roughly solar convection strength with a slight increase from the red giant phase onward. We discover that the convective blueshift shows a local minimum for subgiant stars, presenting a sweet spot for exoplanet searches around higher mass stars, by taking advantage of their spin-down during the subgiant transition.}
  {}


   \maketitle
%

\section{Introduction}
\label{sec:intro}
Convection in stars that have evolved off the main sequence (MS), namely subgiants (SG), red giants (RG), and stars on the asymptotic giant branch (AGB), is a very important factor in their further evolution as well as for their observed properties and overall observability. That is because, as a star leaves the MS, it strongly increases in radius and so does the proportion of the star that is part of the convection zone \citepads{2017A&ARv..25....1H}. Consequently, understanding the structure, strength, and evolution of convective motion in evolved stars is of very high importance to understand post-MS stellar evolution as a whole. Stellar structure simulations predict strong changes in the relative convection zone depth for later evolutionary stages, leading to overall deeper elemental mixing and eventually the so-called dredge-ups where freshly fused elements get transported all the way to the surface. Furthermore, the convective surface structure changes fundamentally (\citeads{2011ApJ...741..119M}, \citeads{2014arXiv1405.7628M}). This can be observed as a deviation from the well-known solar granulation pattern of thousands or millions of small granules for MS stars to only a few, massive granules (\citeads{1975ApJ...195..137S}, \citeads{2013ApJ...769...18T}), potentially spanning over a quarter of the visible surface on red giants \citepads{2018Natur.553..310P}. Observationally, this is of high importance because convective granulation leads to the phenomenon of convective blueshift \citepads[CBS,]{1987A&A...172..200D}. This is a result of the flux imbalance between large, hot granules and smaller, cooler intergranular lanes. The larger influence of the upward moving material inside granules due to their higher spatial and flux weighing overcompensates for the contribution of the intergranular lanes. Projected onto the line of sight toward the observer and integrated over the visible stellar disk, this results in an overall blueshift of the observed spectral lines. The strength of CBS varies with time, as the granulation pattern evolves and can differ significantly between stars that have different convection strengths and structures.\par

Besides the obvious reasons for there being interest in post-MS stars from stellar evolution and structure research, they are of interest to exoplanet researchers as well \citepads{2022AJ....163...53S}. For one, the distribution of planets and their orbits around evolved stars provides insight into the fate of planets when compared to their younger MS counterparts. From the observed differences, it is possible to infer details as to the orbital decay due to tidal forces and improve the models for planetary system evolution (\citeads{2009ApJ...705L..81V}, \citeads{2018ApJ...861L...5G}). It is also expected that the key to the discovery of the mechanism behind inflated gas giants may be found within evolved stellar systems \citepads{2016ApJ...818....4L}, because the most prominent explanations give distinguishably different predictions primarily in these surroundings (reinflation: \citeads{2017AJ....154..254G} vs. delayed cooling: \citeads{2010ApJ...714L.238B}; to give just two examples). Lastly, the spin-down experienced during the subgiant phase significantly lowers the projected rotational velocity of the evolved stars. This allows one to hunt for planets around stars whose fast rotation during the MS phase inhibits or outright prohibits the detection due to rotational spectral broadening and high activity levels with spectroscopic methods \citepads{2013ApJ...774L...2L}. The question about the relation between the planet occurrence rate and host star mass, which appears to show a correlation \citepads{2010ApJ...709..396B}, heavily depends on this.\par

Unfortunately, not many planets have been found to date around evolved stars, accounting for only $\sim$3\% of all exoplanet candidates (150 \citetads{2018PASJ...70...59T}, 135 \citetads{2021ApJS..256...10D}; out of 4900 confirmed planets\footnote{From the \url{exoplanet.eu} catalog}). Two reasons for this are the strong focus on exoplanet searches around solar-like rather than evolved stars and the difficulties involved with high-precision studies of giant stars. While stellar activity and related jitter decreases from spin-down during the subgiant phase, convection, granulation, and related effects become much stronger due to the fundamental changes in stellar structure (\citeads{2020AJ....159..235L}, \citeads{2013Natur.500..427B}). Similarly, stellar oscillations become much stronger over the post-MS evolution as predicted by the scaling relations from\citetads{1995A&A...293...87K}, among others. From an amplitude of $\sim$20\,cm\,s$^{-1}$ and a period of $\sim$5\,minutes for the Sun this can increase to tens of meters per second and multiple days for highly evolved stars, making it a much bigger obstacle than for MS stars \citepads{2019AJ....157..163C}.\par
Besides that complication, \citetads{2016MNRAS.457.3637H} have already classed granulation on its own as the biggest problem for MS planet searches. Its effects can not be mitigated sufficiently through longer integration time or coaddition, can not be recovered through photometric-spectroscopic synergies, or correlations with spectroscopic activity indicators. The later is because the correlations between the classical indicators and convective blueshift variations are too weak (see their Fig. 10 for examples). Instead \citetads{2016MNRAS.457.3637H} suggest the use of the unsigned magnetic flux, which they derived from SDO HMI magnetogram images, as a more direct proxy of the CBS variability. As this quantity is very hard to determine for other stars with a reasonable degree of confidence, and since one is interested in changes in CBS itself, a better approach might be to measure those changes directly and thereby circumvent the need for well correlated proxies. In our previous paper, \citetads{2021A&A...654A.168L}, we describe a method to do this in a robust and purely empirical way. Using the described technique, we find that CBS in early G-type dwarfs like our Sun is roughly 4-5 times stronger than for mid-K types and increases with the third power of the stellar effective temperature after a plateau for late K-dwarfs. This leads to a matching strength increase for CBS related effects that currently prevent the detection of true Earth-twins \citepads{2015A&A...583A.118M} and limit our detection capabilities for Earth-mass planets to late type stars. For evolved stars the problem is even more extreme due to their fewer, larger granules, increased contrast to the intergranular lanes \citepads{2013ApJ...769...18T}, and more massive starspots that inhibit convective motion. All this comes together to form a "flicker floor" that increases with decreasing surface gravity \citepads{2013Natur.500..427B}. Therefore, a better understanding of convective granulation on evolved stars on an empirical level is paramount to the resolution of all these questions, even more so than for MS stars.\par
Solving this problem is especially important in the face of the massive amounts of data that have become available from past and current space missions, for instance Kepler/K2 and TESS, and will become available from future missions such as PLATO. While not dedicated toward evolved stars, the observation fields still contain tens of thousands of them that we can not exploit until our current inability to properly account for their intrinsic variability has improved.\par
With this motivation, we seek to measure the CBS strength for post-MS stars in the subgiant, red giant, red clump or horizontal branch, and asymptotic giant branch phases and thereby provide empirical convection strengths, as we did in \citetads{2021A&A...654A.168L} for MS stars. 

\section{Data and Processing}
In this paper we extend our work on a sample of HARPS spectra from \citetads{2021A&A...654A.168L}, called paper I from here on, from the main sequence to the previously excluded post-MS regime. To that end, we apply the technique from paper I, summarized in Sect. \ref{sec:technique}, to the previously excluded sample of stars and supplement with a sample of spectra from the PEPSI spectrograph for comparison. Further, we determine the evolutionary stages of the post MS stars from the sample to better interpret the results.

\subsection{Observations}
\begin{figure}
\resizebox{\hsize}{!}{\includegraphics{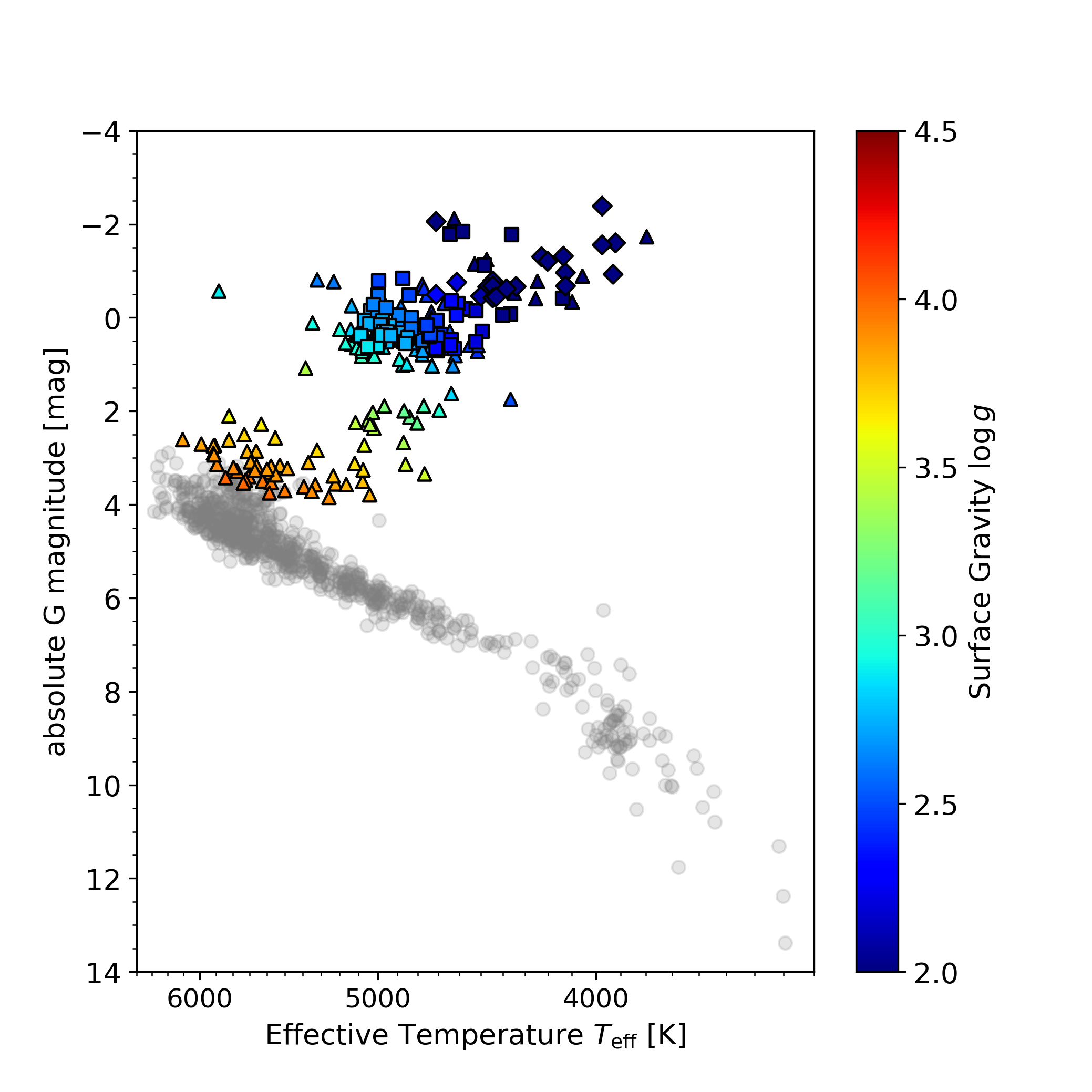}}
\caption{Our sample of HARPS stars (post-MS colored for surface gravity fitted from MIST model, MS stars from paper I in gray) in an Hertzsprung-Russel diagram, based on \textit{Gaia} DR2 data. The marker shapes correspond to the evolutionary phases according to MIST: main sequence (circle), sub- or red giant (triangle), core Helium burning (squares), and asymptotic giant (diamonds).}
\label{img:HR}
\end{figure}

The primary data set behind this paper and paper I was compiled by \citetads{2020A&A...636A..74T}, who collected, sorted, and filtered all publicly available spectra from the HARPS\footnote{High-Accuracy Radial velocity Planetary Searcher} spectrograph, a fiber-fed, cross-dispersed echelle spectrograph with a resolving power of R=115\,000. HARPS spans the range between 380\,nm and 690\,nm over 72 echelle orders. It is housed in a temperature stabilized, evacuated chamber at the 3.6\,m telescope at La Silla, Chile and is capable of reaching a precision of 1\,m\,s$^{-1}$ \citepads{2003Msngr.114...20M}. \citetads{2020A&A...636A..74T} coadded all available spectra that belong to the same object with the SpEctrum Radial Velocity AnaLyser code (\texttt{serval}, \citeads{2018A&A...609A..12Z}) which also provides radial velocities (RVs) from matching the coadded spectra as templates to the individual observations. Nightly offsets and long-term trends, for example binary motion, were corrected for the individual spectra as well. We make use of the coadded template spectra, after converting them to air wavelengths, for their increased S/N as well as referenced the RV values listed in the \emph{RVBANK} from \citetads{2020A&A...636A..74T}. The individual, coadded spectra were processed, normalized, and analyzed identically to paper I (see Sect. \ref{sec:technique} for a summary, paper I for details). Figure \ref{img:HR} shows the MS sample from paper I for comparison together with the post-MS sample analyzed in this paper in a Hertzsprung-Russel (HR) diagram based on \textit{Gaia} DR2 parameters and MIST synthetic photometric fits (Sect. \ref{subsec:stellar_evo}). It was decided to stay with DR2 parameters instead of adopting the newly released (E)DR3 catalog values to keep the results consistent with paper I.\par
The S/N ratios provided by \texttt{serval} for the coadded spectra are shown in Fig. \ref{img:data_S/N_hist} in the top panels for each star individually (left) and binned by S/N (right, stacks colored to reflect evolutionary phase). On average, 27 spectra were used per coaddition for an approximate S/N of 555, with some stars significantly higher. The sample, binned by temperature and color-stacked by evolutionary phase, is shown in the lower panel of Fig. \ref{img:data_S/N_hist}. Table \ref{tab:star_data} gives a complete list of the stars from the final sample and their parameters.\par
In addition, we supplement our sample with data from the PEPSI\footnote{Potsdam Echelle Polarimetric and Spectroscopic Instrument} spectrograph, published by \citetads{2018A&A...612A..45S}, for comparison and verification of our results from HARPS (Appx. \ref{apdx:gray_investigate}). PEPSI is a fiber-fed instrument mounted at the 2\,x\,8.4\,m Large Binocular Telescope (LBT) on Mt. Graham, Arizona, USA that covers the range of 383 - 907\,nm at a resolving power of up to 270\,000 \citepads{2015AN....336..324S}. We were able to use the data as provided for our line-by-line measurements, as the spectra are corrected for RV, continuum normalized, stitched, and given in air wavelengths, with no further preparation necessary. The data set is visualized in an HR diagram (HRD) in Fig. \ref{img:PEPSI_HR}.

\subsection{Data filtering}
\label{subsec:filter_phase}
Out of the complete, 3094 star sample that covers all evolutionary stages, 439 were excluded in paper I for lack of a match with \textit{Gaia} DR2 stars \citepads{2018yCat.1345....0G} which were used as a source of consistent stellar parameters, such as effective temperature values. From the remainder, 810 MS stars were identified, based on SIMBAD\footnote{\url{http://simbad.u-strasbg.fr/simbad/}} surface gravities $\log g$ and \emph{Gaia} $G$ magnitude in conjunction with effective temperature $T_\mathrm{eff}$, and analyzed in paper I. The database values for $\log g$ were used only for this filtering step, further analyses use the values determined in Sect. \ref{subsec:stellar_evo}. This paper investigates the 458 stars that are not included in paper I as they were presumed to be post MS. In paper I we find that we need to exclude stars with a projected rotational velocity of $v \sin i >$ 8\;km/s due to the effect the combined instrumental and rotational broadening has on our analysis technique, which is summarized in Sect. \ref{sec:technique}. Our $v \sin i$ values are taken primarily from SIMBAD, followed by \citetads{2005yCat.3244....0G}. As a proxy, we further use the HARPS Data Reduction Systems' (DRS) Full-Width Half-Maximum (FWHM) values from \citetads{2020A&A...636A..74T}. They do not correlate perfectly with $v \sin i$ but provide a first order approximation that allows to exclude a few more fast rotating stars that do not have velocities listed in the other references.\par
For the PEPSI data, despite the higher resolving power of the instrument compared to HARPS, we have retained the 8\,km\,s$^{-1}$ $v \sin i$ limit because the decrease in instrumental broadening only allows a negligible extension of 200\,m\,s$^{-1}$ to the acceptable projected rotational velocity. 

\subsection{Stellar evolution phase and surface gravity}
\label{subsec:stellar_evo}
The preliminary sample for this work, after the $v \sin i$ and FWHM filtering of the presumed post-MS sample, comprises 267 stars. The specific phases in the stars' evolution, as well as surface gravities, were determined from matching synthetic photometric observations to a set of \textit{Gaia} DR2 parameters. We use the MESA Isochrones \& Stellar Tracks (MIST; \citeads{2016ApJS..222....8D}; \citeads{2016ApJ...823..102C}; \citeads{2011ApJS..192....3P}, \citeyearads{2013ApJS..208....4P}, \citeyearads{2015ApJS..220...15P}) projects synthetic data to determine the evolutionary phase in accordance with their classification. MIST does not feature subgiants as a distinct group but they can easily be distinguished from red giants through surface gravity. The phase was determined with the UBV(RI)c + 2MASS JHKs + Kepler + Hipparcos + Tycho + Gaia, version 1.2 grid of synthetic photometric data created for $\left[\mathrm{Fe/H}\right]=0.0$, $v/v_\mathrm{crit}=0.4$. We matched the photometric grid for the effective temperature, luminosity, and G, BP, and RP magnitudes from \emph{Gaia} DR2 of our sample to obtain the grid point that best matches all these parameters. Since this only matches the grid points to the known parameters without interpolating the evolution tracks, we can not define uncertainties on the determined best matches.\par
Instead, for the surface gravity log g, we defined an error box based on the \textit{Gaia} uncertainties in effective temperature and luminosity and collected the surface gravities of all MIST grid points within the error box of each star. The minimum and maximum within each box were then taken as the uncertainty boundaries of the matched surface gravity. The mean calculated over the maximum deviation between those boundaries and the matched surface gravities gives an average uncertainty of 0.16\;dex. The comparison of that uncertainty estimate to the literature via a SIMABD query with \texttt{Astroquery} shows a good agreement with no systematic deviations between the values (mean over the signed deviations is 0.002\;dex). The mean over the absolute deviations between the two data sets is 0.25\;dex, consistent to our error box derivation. The deviations between our results and the literature compared to the spread within the literature show similar magnitudes, indicating that the algorithm performs as intended.\par
\begin{figure*}
\resizebox{\hsize}{!}{\includegraphics{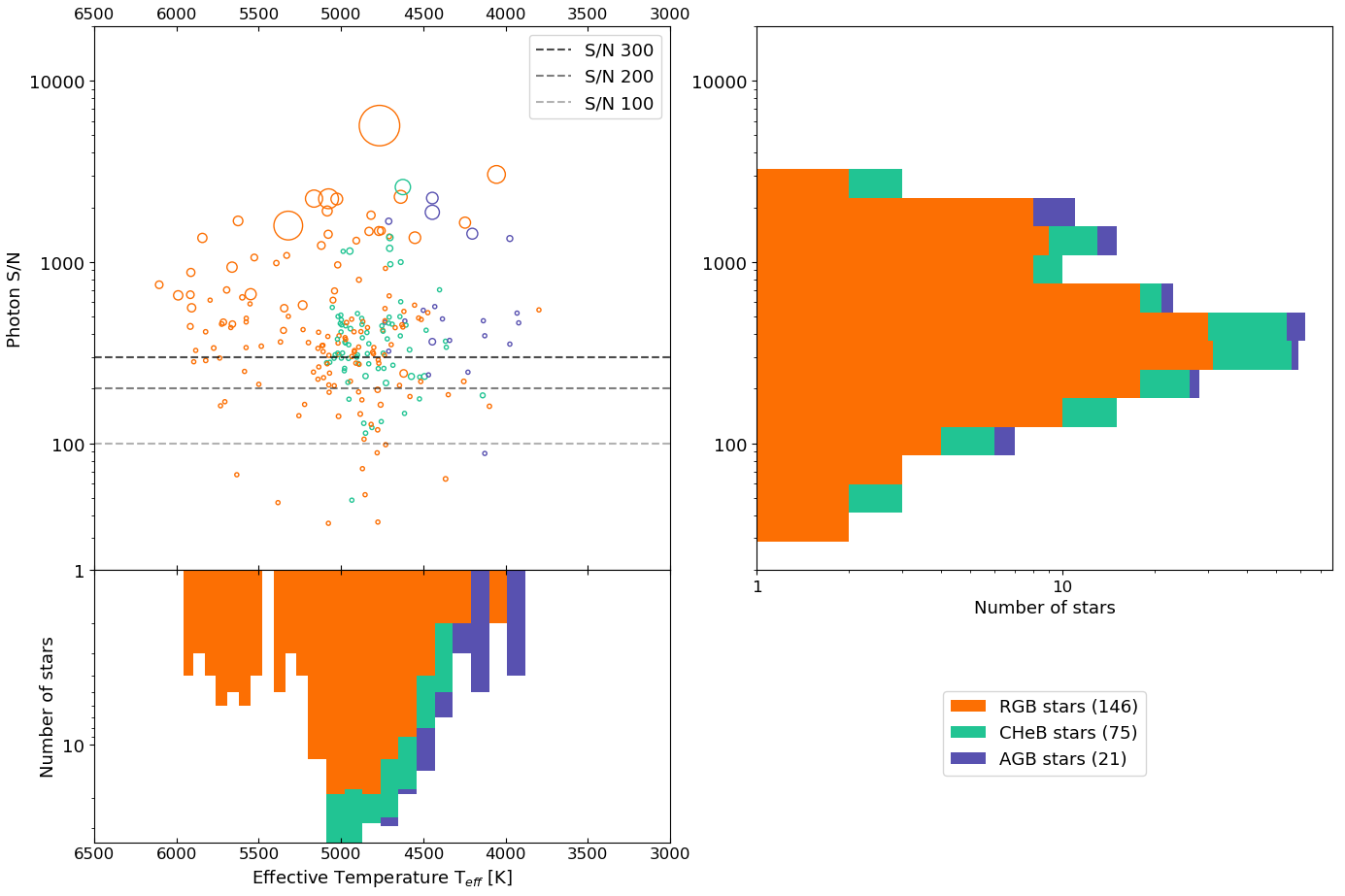}}
\caption{Overview of all final sample stars' signal-to-noise ratio (S/N) from \texttt{serval}, temperature from \textit{Gaia} DR2, and evolutionary phase according to MIST photometry (color coded; the legend shows the number of stars per phase). Top left: Distribution of S/N over temperature within the sample. The circle size corresponds to the number of observations for that star. The dashed, horizontal lines mark S/N values of 100, 200, and 300. Top right: S/N for the coadded spectra in the sample. Bottom left: Distribution of temperatures within the sample.}
\label{img:data_S/N_hist}
\end{figure*}

Besides the surface gravities of the underlying stellar models and the classification of the evolutionary stage, the MIST matches further reveal a set of 25 stars that were excluded in paper I based on low surface gravity values that appear to be incorrect. They are identified from the MIST grid as MS stars and are excluded from the 267 remaining post-MS stars. This leaves a set of 242 post-MS stars that span subgiants, red giants, horizontal branch, and asymptotic giant branch stars. It is this sample of stars that is shown in Figs. \ref{img:HR} and \ref{img:data_S/N_hist}. Closer review of the results in Sect. \ref{sec:results} excluded one more star, HD87833, from the set. It has no rotational velocity listed in our two primary sources and the FWHM is acceptable, though \citetads{2013A&A...554A...2S} list it at a $v \sin i$ of 8\,km\,s$^{-1}$, which removes it from our sample. This leaves a final, trusted sample of 242 post-MS stars.

\section{Technique}
\label{sec:technique}
To determine the strength of convective blueshift for the stars in the sample, we follow the third-signature scaling approach \citepads{2009ApJ...697.1032G}. In the following, we provide a short summary of the technique employed here and in paper I. For a more in-depth explanation we refer to Sects. 5 and 6 from paper I.\par
The core of the method is the concept of the third signature of granulation, the relation between a spectral lines' absorption depth and its radial velocity shift through convective motion \citepads{1999PASP..111.1132H}. It can be used as an alternative to the classical bisector approach, where connected midpoints of a single spectral line along points of equal absorption depth in the red and blue wing are used to trace convective velocity through the photosphere. The third-signature technique replaces the single line sampled at many depth points with many lines sampled at a single depth point each: Their core. This increases the number of required lines from one into the hundreds, but decreases the required signal-to-noise from S/N$\gtrsim$300 to $\approx$100 and resolving power from R$\gtrsim$300\,000 to R$\approx$50\,000 as fitting a line core is much more robust than interpolating a bisector from the wings. This is especially true on the higher (close to the continuum) and lower (close to the core) end of the line profile due to noise combined with a small spectral slope. The lowered requirements, especially in resolving power, allow for the use of a much larger set of broad-band instruments that more than compensates the increase in line numbers necessary.\par
\begin{figure}
\resizebox{\hsize}{!}{\includegraphics{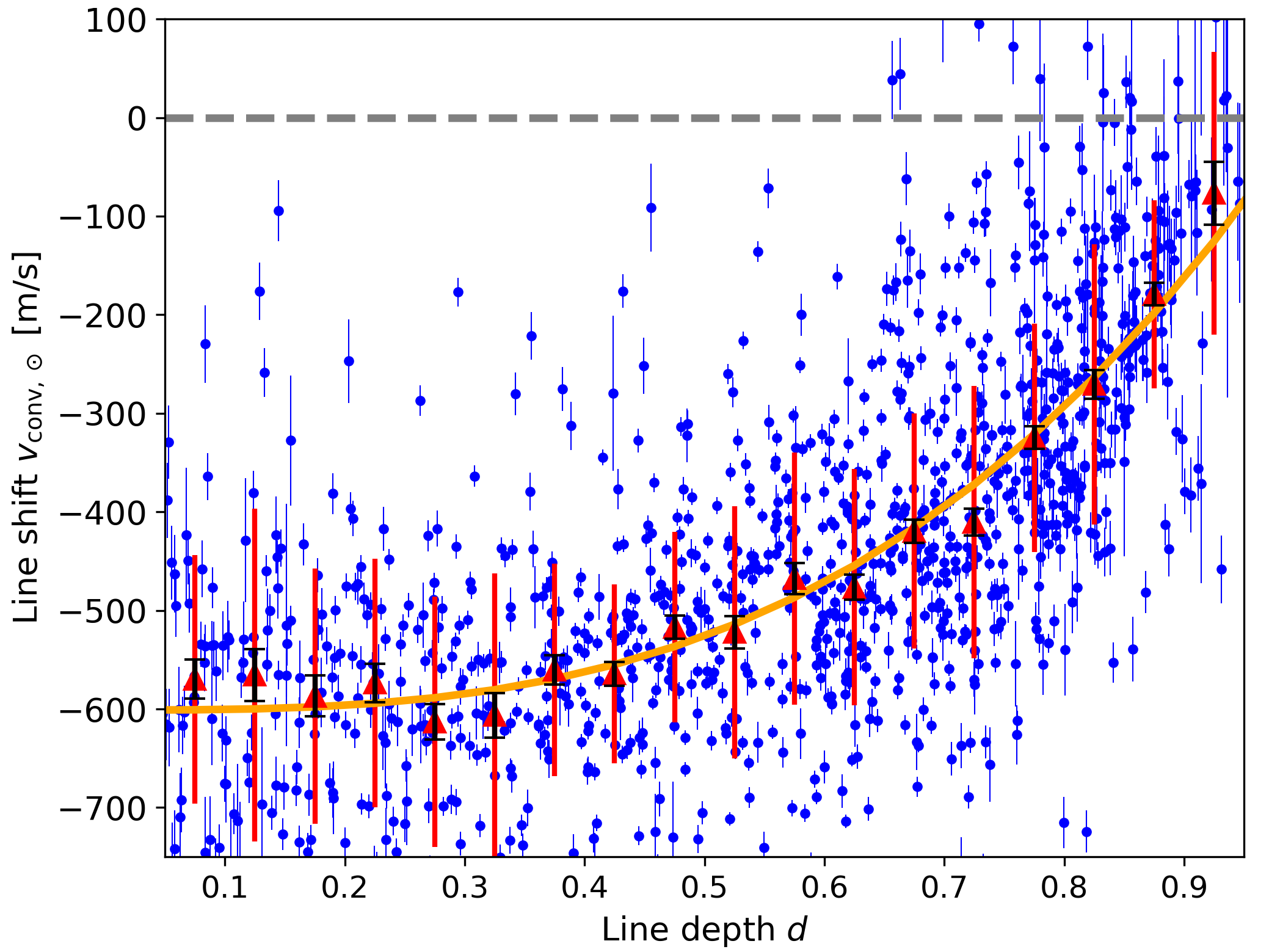}}
\caption{Line shift vs. line depth for the solar template spectrum and the third signature of granulation template used in this work. This plot is modified from paper I, Fig. 5, center-left panel, with adjusted axes but using the same data, binning, and fitted signature. Marked are the 1168 individual spectral line measurements from the VALD list (blue points with errorbars), 18 bin medians (red triangles, the red bars indicate the median absolute deviation), error of the bin median (black errorbars, analogous to the error of the mean), and the template third signature (orange line, Eq. (\ref{eqn:thirdSigModel})).}
\label{img:3rdsigexmpl}
\end{figure}

An additional advantage of the third-signature technique compared to the bisector approach is its universal nature. Unlike bisectors, which differ significantly even between lines of the same species in the same spectrum, the third signature of any star below the granulation boundary (\citeads{2010ApJ...721..670G}; $\sim$7000\;K on the MS; cooler for giants) can be scaled and shifted in velocity with a scale factor $S$ and an offset to match any other star below the granulation boundary \citepads{2009ApJ...697.1032G}. This allows for the use of a high quality template star, the Sun in our case, to fit a model for its third signature and then calibrate and scale that template model to other instruments and stars, significantly improving the robustness of CBS determination. The calibration step is required to account for changes in instrumental broadening. In this work we use the same template as in paper I for the solar blueshift velocity $v_\mathrm{conv, \odot}$ as it depends on the line absorption depth $d$:
\begin{align}
v_\mathrm{conv, \odot}\left(d\right) = 601.110\;\mathrm{ms}^{-1} \cdot d^3 + 173.668\;\mathrm{ms}^{-1}.\label{eqn:thirdSigModel}
\end{align}
The template is based on the IAG solar flux atlas \citepads{2016A&A...587A..65R} with a resolving power of $R\sim 1\,000\,000$ and created from a filtered list of spectral lines from the \emph{Vienna Atomic Line Database\footnote{\url{vald.astro.uu.se}}} (VALD; \citeads{1995A&AS..112..525P}, \citeads{2000BaltA...9..590K}, \citeads{2015PhyS...90e4005R}; For details see paper I, Sect. 4.2). It is shown in Fig. \ref{img:3rdsigexmpl} and includes the (median binned\footnote{The astropy function mad\_std was used to calculate the median absolute deviation (MAD) to quantify the bin variance. This function also scales the MAD to be consistent with the sample standard deviation}) line measurements underlying the fit from the solar spectrum, but is vertically shifted by 775\,m\,s$^{-1}$ such that $v_\mathrm{conv, \odot}\left(1.0\right) = 0$ to account for gravitational redshift and the motion of the Earth and telescope. The choice of vertical zero point is arbitrary as CBS is purely differential, without an absolute component. The template further needs calibration for the HARPS resolving power by 0.91 in scale and 21.2\;ms$^{-1}$ in offset to account for differences in instrumental broadening. These are implicitly applied at all points for the rest of this paper.\par
A big advantage of obtaining CBS strength through third signature scaling instead of for example using the bisector inverse slope or other bisector indicators is that, in addition to the global CBS strength, it provides a velocity profile over all possible line depths. This profile allows for the construction of CBS values for the core of any spectral line of known depth even outside the initial list of measured lines. It further allows for the reconstruction of the lower parts of the bisector, where the intergranular contribution is weak, and serves as a proxy to the convective velocity profile within the range of optical depths that belong to the line forming region within the stellar photosphere.

For the rest of this paper we use the same curated VALD line list used in paper I to extract the third signature. All analyses, unless stated otherwise, are carried out the same way as those from paper I. We started with the normalization of all spectra to the continuum by performing a boundary fit with asymmetric weighting and clipping factors, following a modified version of \citetads{2009MNRAS.396..680C} and using a sinc$^2$ function to simultaneously account for the instrumental blaze function. We calculated an approximate RV from the cross-correlation of the normalized spectra with a binary line mask from CERES \citepads{2017PASP..129c4002B} by averaging over the echelle orders and determining the minimum as an initial RV estimate. We then fitted the line list, determined in paper I, to the spectra, using parabolas to approximate the line core. We repeated the fit until convergence was reached, starting again from the supposed line center determined from the previous step to account for errors in the initial RV guess. After fitting the solar third signature template to the measured line center and depth values, the RV from the cross-correlation function (CCF) was refined such that $v_\mathrm{conv}\left(1.0\right)=0$. This places a theoretical, fully absorbing line at zero RV shift under the assumption that it forms at the very top of the convection zone and should, therefore, show no convective velocity. Any deviation from zero would then necessarily be due to uncorrected RV. Measuring the lines and refining the RV was repeated until convergence to the final RV. For the full details we refer to paper I, Sects. 4 and 6.

\section{Results}
\label{sec:results}
\begin{figure*}
\resizebox{\hsize}{!}{\includegraphics{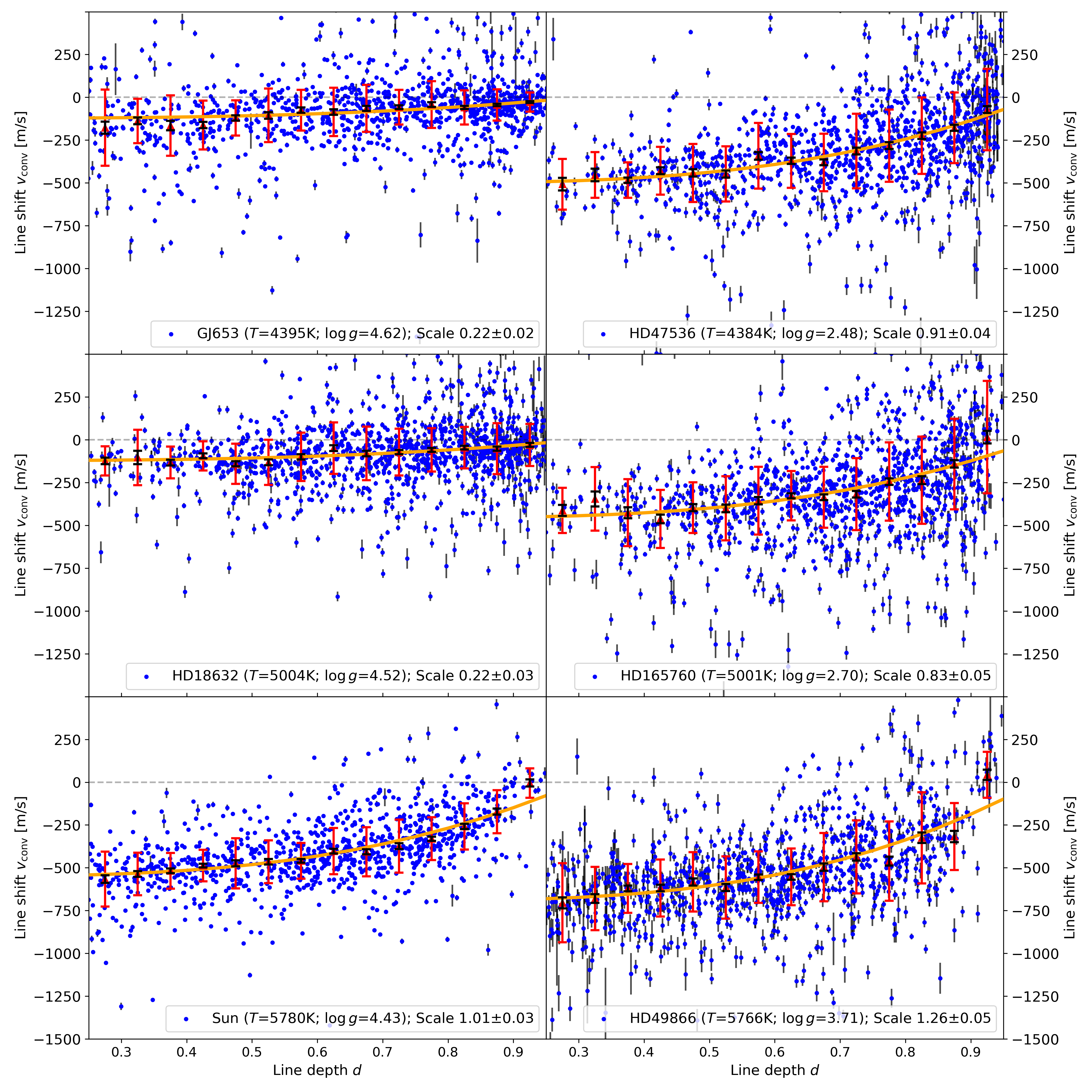}}
\caption{Comparison of the convective blueshift of selected MS (left) and post-MS (right) stars of similar effective temperatures (rows) from the HARPS samples. The more advanced evolutionary stage can be inferred from the significantly decreased surface gravity log\,$g$ for the stars on the right. Marked are individual spectral line measurements from the VALD list (blue points with errorbars), bin medians (red triangles, the red bars indicate the median absolute deviation), error of the bin median (black errorbars, analogous to the error of the mean), and the fitted third signature (orange line, scale factor in legend).}
\label{img:3x2_example}
\end{figure*}
Applying the third signature scaling technique as described in Sect. \ref{sec:technique} to MS stars in paper I revealed a strong correlation between the strength of CBS and effective temperature that monotonously increases from a 20\% solar strength plateau for late K dwarfs to 150\% for mid to late F dwarfs. Figure \ref{img:3x2_example} exemplifies the general results for the post-MS sample of this work compared to paper I. In a direct comparison between three post-MS stars, selected from this works data set, and three MS stars of near identical temperature from paper I, we see a significant increase in CBS strength for the post-MS stars compared to their less evolved MS counterparts. The difference between the post-MS stars and their counterparts also appears to grow with decreasing surface gravity, meaning for advancing evolution. The absolute CBS strength appears to dip for the intermediate case.

\subsection{Temperature dependence of post-MS CBS strengths}
\label{subsec:temp_dep}
\begin{figure*}
\resizebox{\hsize}{!}{\includegraphics{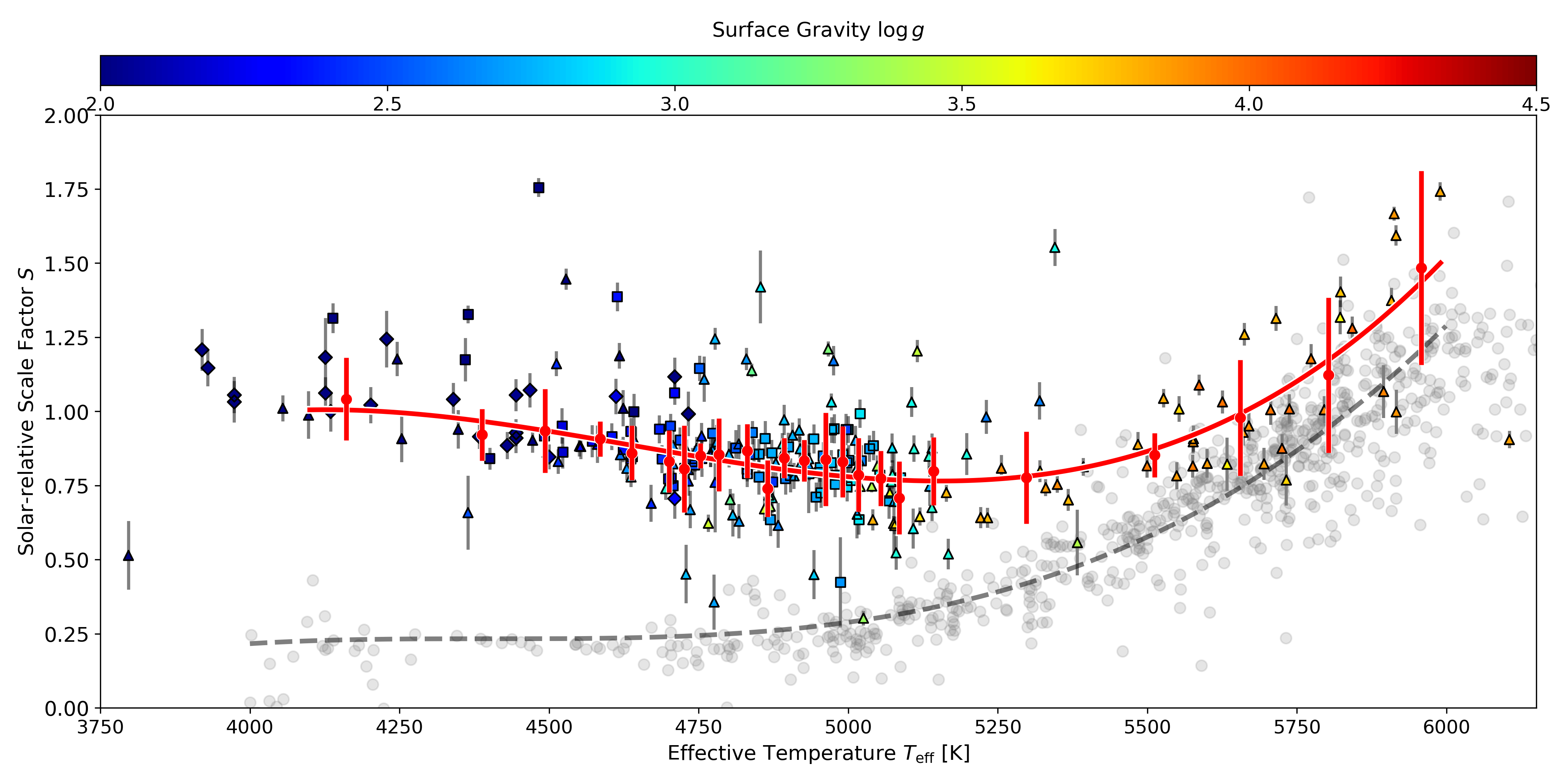}}
\caption{Scale factor vs. effective temperature for the post-MS stars of the HARPS sample. The relation for the post-MS stars is fitted with a cubic polynomial (red line, Eq. (\ref{eqn:SG_fit})) to the binned scale factors (red bars indicate the median absolute deviation), while the MS results (gray, dashed line for the fitted relation $S_\mathrm{HARPS,MS}\left(T_\mathrm{eff}\right)$, gray markers for individual stars hotter than 4000\,K; paper I) are included as reference. The surface gravity for each post-MS star (based on MIST fits) is color coded. The marker shapes correspond to the evolutionary phases according to MIST: sub- or red giant (triangle), core Helium burning (squares), and asymptotic giant (diamonds).}
\label{img:scalesplot}
\end{figure*}
The results from the full post-MS sample are given in Fig. \ref{img:scalesplot}. They match Fig. \ref{img:3x2_example} and show the gradual deviation of the post-MS stars from the MS relation discovered in paper I. While the MIST stellar evolution grids do not list subgiants as their own stage, they are still easily identifiable through their surface gravity. Subgiants make up the transition between the MS relation and the post-MS branch between 5300\,K and $\sim$6000\,K, lead smoothly into the red giant branch (RGB) stars and terminate at AGB stars after passing horizontal branch stars. The entire post-MS sample forms its own "horizontal branch" in scale factor - effective temperature space at roughly solar strength CBS. We approximate the observed relation between Scale factor $S$ and effective temperature $T_\mathrm{eff}$ with a third order polynomial to bins of ten stars each, same as in paper I for the MS sample:
\begin{align}
t &= \frac{T_\mathrm{eff} - 4400\,\mathrm{K}}{1000\,\mathrm{K}},\\
S\left(t\right) &= 0.963 - 0.275\cdot t - 0.317\cdot t^2 + 0.442\cdot t^3.\label{eqn:SG_fit}
\end{align}
The effective temperature of the star is scaled in accordance with paper I and the overall relation is fitted for the temperature region of 4100\;K - 6000\;K. Above 5800\;K the two relations, for MS and post MS, fall together within the margin of error for terminal age main sequence (TAMS) stars, in accordance to the turnoff within the HRD in Fig. \ref{img:HR}. This is also shown in Fig. \ref{img:CBS_HR} where the HRD is color coded for CBS strength. An overview of the CBS strengths expected from the fit is given in Table \ref{tab:cheatsheet} for a selection of effective temperatures within the fitted range.\par
We evaluate the robustness of this fit in several ways. First, a Kolmogorov-Smirnov test was performed of the fit residuals against a standard normal distribution. In the case of an accurate fit, the two should be identical with over- and underfitting apparent by a narrower or wider residual distribution relative to the standard normal. This shows that, while a third order polynomial does, technically, overfit the binned data, a second order polynomial strongly underfits the unbinned data compared to third order and visually deviates for low and high temperatures. The technical overfit is further to be expected from the choice to directly use the median absolute deviation as a bin uncertainty measure without scaling by the square root of the number of bin members as was done for Figs. \ref{img:3rdsigexmpl} and \ref{img:3x2_example}. This is motivated by the much smaller number of bin members and stronger effect of systematics such as activity that count against the assumption of similarity between the mean and median estimators employed previously. Therefore, the third order polynomial is considered the better choice.\par
Secondly, the impact of the uncertainty of the \textit{Gaia} temperature on our fit is investigated with orthogonal distance regression (ODR) which generalizes the residual definition of the standard chi-square minimization to multiple dimensions. Since the \textit{Gaia} uncertainties are asymmetric because they are defined via upper and lower percentiles, the uncertainty interval also has to be symmetrized. For this we took the minimum, maximum, and mean of the upper and lower uncertainty as a symmetrical uncertainty for the ODR. The results from the ODR performed on either the binned data, now also with the MAD included as bin variance in temperature, or directly on the data with symmetrized uncertainties in temperature are in agreement with the original fit shown in Fig. \ref{img:scalesplot} within the region of error in all cases. The effect that the symmetrized uncertainties might have is unknown as we are unaware of a fitting suite that can correctly handle such a case. It appears plausible that the generally larger lower uncertainty could, in the case of proper handling, introduce an effective offset in the fit that can not be caught with symmetrical errors though we are unable to confirm or deny this.

\subsection{Surface gravity dependence of post-MS CBS strengths}
\begin{figure*}
\resizebox{\hsize}{!}{\includegraphics{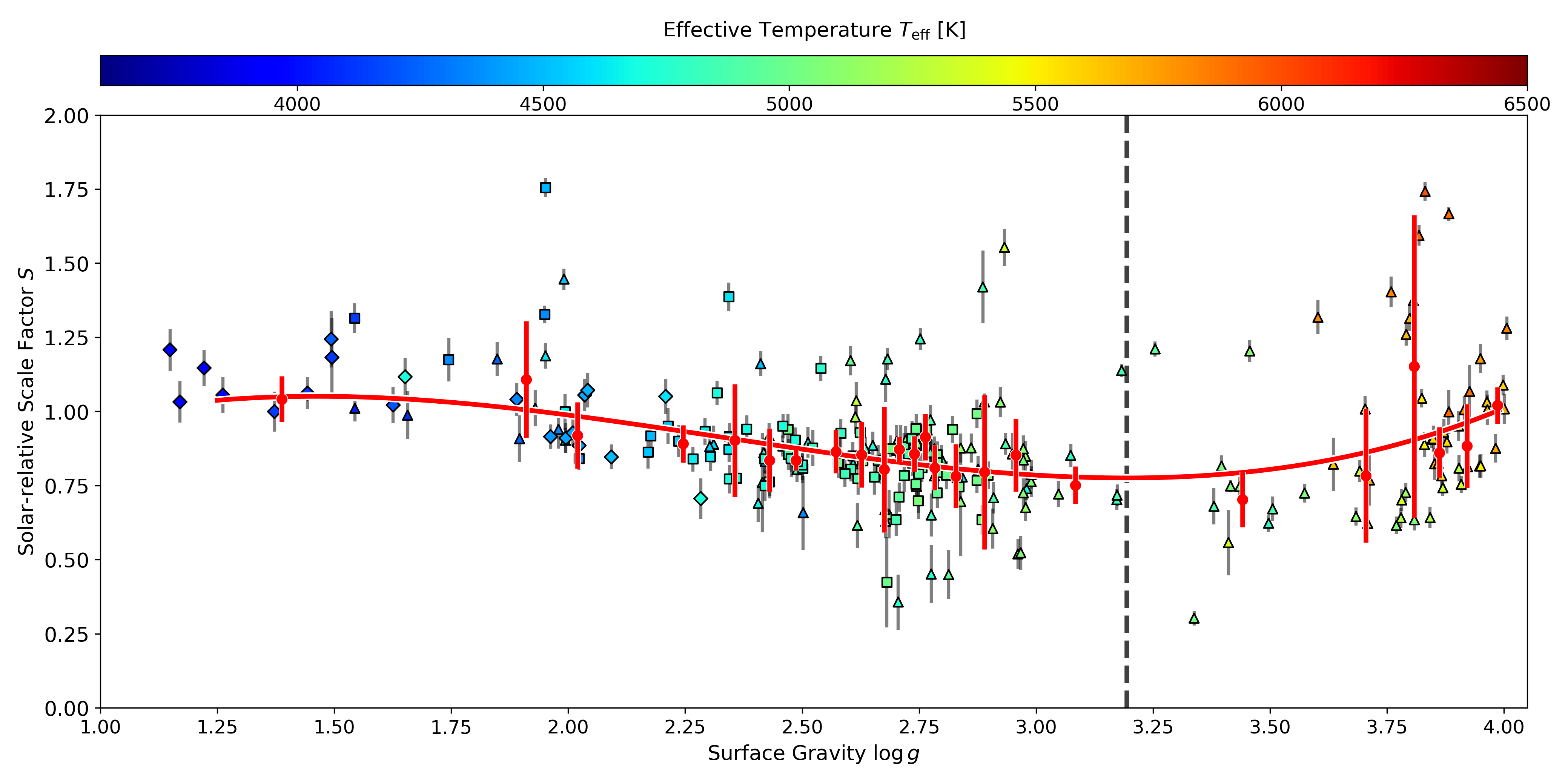}}
\caption{Scale factor vs. surface gravity (based on MIST fits) for the post-MS stars of the HARPS sample. The relation $S_\mathrm{HARPS,pMS}\left(\log g\right)$ for the post-MS stars is fitted with a cubic polynomial (red line). The effective temperature for each star is color coded. The marker shapes correspond to the evolutionary phases according to MIST: sub- or red giant (triangle), core Helium burning (squares), and asymptotic giant (diamonds). A local minimum in scale factor at log\,$g\approx3.2$ is marked (gray, dashed line).}
\label{img:S_vs_logg}
\end{figure*}
The plot of the derived scale factors $S$ against the surface gravity $\log\,g$ instead of temperature shows a similar picture due to the dependence of the two but also clearly shows an intermediate decrease in scale factor for the youngest subgiants with a minimum around $\log\,g \approx 3.2$ (Fig. \ref{img:S_vs_logg}). Fitting the relation with a third order polynomial, as with eqn. (\ref{eqn:SG_fit}), gives:
\begin{align}
S\left(\log g\right) = 0.129 + 1.488\log g - 0.742\left(\log g\right)^2 + 0.106\left(\log g\right)^3.\label{eqn:SG_fit_logg}
\end{align}
As within Sect. \ref{subsec:temp_dep} for the effective temperature fit, we again investigate the robustness of the surface gravity fit. For this we use the uncertainties derived in Sect. \ref{subsec:stellar_evo}. Like with the temperature, we first symmetrized the uncertainty interval and then fitted the polynomial against the binned data with the MAD as bin variance and directly against the data that was symmetrized to the minimum, maximum, and mean of the asymmetric errorbars. The results are very similar to the previous ones for the temperature, with only minor changes to the fitted parameters and the polynomial behaves the same within the uncertainty of the data.\par
The intermediate dip in $S$, present in all fitting approaches at the same surface gravity, is in agreement with results from \citetads{2020AJ....159..235L}, who investigated RV jitter levels due to activity and granulation. They also find a minimum in RV jitter which depends on surface gravity and stellar mass. It separates what they call an activity dominated and a granulation and oscillation dominated phase during a stars evolution. For a median mass of 1.78 solar masses in our sample they predict a jitter minimum at $\log\,g \approx 3.1$ of 5.5\,m\,s$^{-1}$ in coincidence with the minimum observed in our results. This could indicate a sweet-spot where granulation and oscillation induced RV jitter as well as CBS are both minimal, reducing the difficulties encountered by high-precision planet searches (Sect. \ref{sec:intro} for details).

\subsection{Comparison with literature results}
\label{subsec:lit_compare}
The results for post-MS CBS strengths are in qualitative agreement with the results from \citetads{2009ApJ...697.1032G}, who also shows an increase in CBS for more evolved stars, however they differ on a quantitative level where the \citetads{2009ApJ...697.1032G} results show significantly higher scale factors for the most evolved stars. We perform a thorough investigation into the source of this discrepancy as described in Appendix \ref{apdx:gray_investigate}. We conclude that it is not a matter of instrumental differences or any errors during the analysis. Instead the discrepancy results from a different, but equally valid, choice of lines which lead to a differently shaped "standard curve" (Gray) or third-signature model (our term) that scales differently with temperature. This is supported by results from \citetads{2012AJ....143..112G} and \citetads{2014ApJ...796...88G} for $\zeta$ Cygni and $\beta$ Gem respectively, both of which are found to show weaker than expected scale factors while using additional lines compared to the \citetads{2009ApJ...697.1032G} reference. As such the results are not in actual disagreement as they are not quantitatively comparable due to the scaling difference of the underlying relation. A similar disparity is found in paper I from the comparison of our results for MS stars against \citetads{2017A&A...597A..52M} and \citetads{2017A&A...607A.124M}. They used a linear approximation as their third-signature model that results in an underestimation of the scaling toward hotter stars due to the strong curvature of the third signature.

\subsection{Comparison with analytical models}
\label{subsec:BC_comapre}
\begin{figure*}
\resizebox{\hsize}{!}{\includegraphics{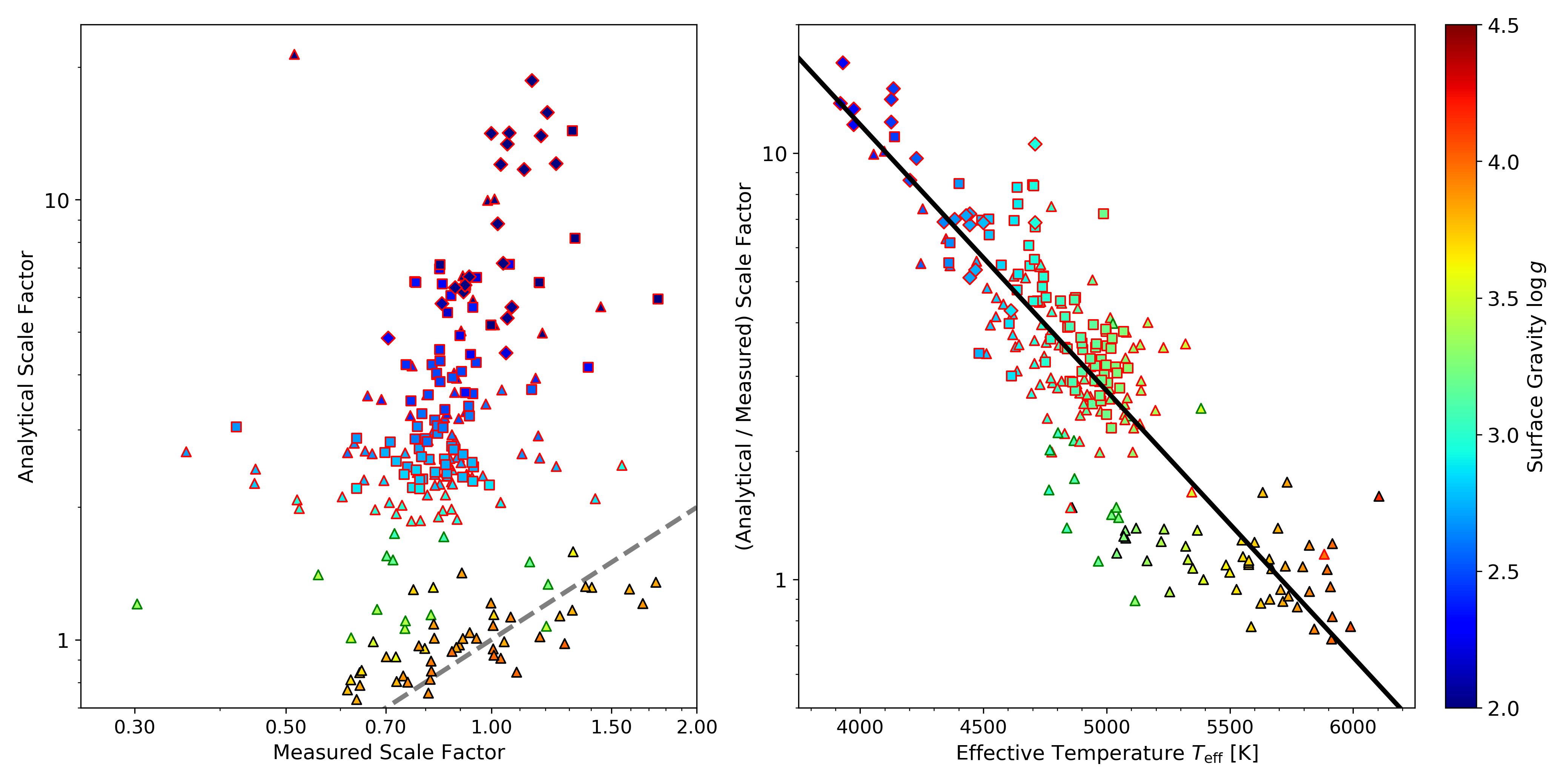}}
\caption{Comparison of analytical and empirical scale factors. Left: Scale factors measured from the HARPS post-MS sample plotted against values expected from the \cite{Basu2017} analytical relations. The gray, dashed line shows equality. Right: Ratio of scale factor values expected from the \cite{Basu2017} analytical relations (Eq. \ref{eqn:BC_CBS}) and measured from the HARPS post-MS sample. The black line was fitted as a correction factor to the analytic values as a function of effective temperature. Both: Markers are colored for surface gravity (based on MIST fits). Black edges indicate $\log g > 3.5$, green $3.5 > \log g > 3.0$, and red $3.0 > \log g$. The marker shapes correspond to the evolutionary phases according to MIST: sub- or red giant (triangle), core Helium burning (squares), and asymptotic giant (diamonds).}
\label{img:BC}
\end{figure*}

While we are able to explain the discrepancy with the results by \citetads{2009ApJ...697.1032G}, its existence leaves the question whether our scale factors are under similar limitations. To that end, following paper I, we again attempt to match our measurements to the analytical scaling relation for the observed CBS $S_\mathrm{BC}$, called $\sigma_\mathrm{c,v}$ by \citet{Basu2017} (Eq. (4.79) in their work), with effective temperature $T$, stellar radius $R$, and stellar mass $M$, shown in Eq. (\ref{eqn:BC_CBS}).
\begin{align}
S_\mathrm{BC}\left(T, R, M\right) \approx T^{41/9} R^{13/9} M^{-11/9}.\label{eqn:BC_CBS}
\end{align}
In paper I, we are able to match our scale factors either for MS stars below or above 5400\,K, though not for the entire range at once as the scaling with temperature for the hotter range appears stronger than analytically expected. The relation also does not reproduce the plateau for K-type MS stars. We conclude that this is due to the underlying assumption by \citet{Basu2017} that individual granules contribute the same level of intensity fluctuations irrespective of spectral type. This impression is reinforced after we compare the current results for post-MS stars to the \cite{Basu2017} relation. With the stellar parameters obtained from the MIST fits (Sect. \ref{subsec:filter_phase}) and radii calculated from the model mass and surface gravity as radius is not a listed parameter, it is possible to match the scale factors for TAMS and young SG stars (Fig. \ref{img:BC}, left panel).

At a surface gravity $\log\,g < 3.5$ the scale factors obtained from the analytical relation start to overestimate our empirical measurements and for $\log\,g < 3.0$ predict a strong increase with further decreasing surface gravity. We only see a very slight increase in CBS scale factor in our data for that region, which corresponds to the transition between SGs and the actual RGB. From the fundamental structural changes experienced during that phase it is unsurprising that the relation breaks under the given assumptions. As with the MS sample, we are of the opinion that this is due to changes in the balance factor between granules and lanes. A comparison of the model from \cite{Basu2017} on the number and therefore size of the granules with numerical results from \citetads{2013ApJ...769...18T} shows a good agreement and reinforces our opinion. Finally, \citetads{2013ApJ...769...18T} directly show that the intensity contrast increases as the star evolves which explains the deviation from the analytical model. Since the deviation of our empirical CBS strengths from the analytical expectation appears to be smooth, we can attempt to derive a correction term. The right panel of Fig. \ref{img:BC} shows the ratio of the analytical and the empirical strength of CBS ($S_\mathrm{BC}$ and $S_\mathrm{HARPS,pMS}$ respectively) for the post-MS (pMS) sample as a linear function of temperature in semilogarithmic space. Fitting the relationship gives Eq. (\ref{eqn:BC_correction_factor}), which can be used to empirically correct the analytically derived values from \cite{Basu2017}.\par
\begin{align}
\log_{10}\left(\frac{S_\mathrm{BC}}{S_\mathrm{HARPS,pMS}}\right) = -0.6244\frac{T_\mathrm{eff}}{1000\,\mathrm{K}} + 3.564.\label{eqn:BC_correction_factor}
\end{align}
One may expect that the point of deviation from the expected behavior lies on the granulation boundary, as delineated in \citetads{1989ApJ...341..421G} as the point in the HRD where the bisectors flip from C to reversed-C shapes. However, the entirety of our sample falls below that formal boundary otherwise the use of a solar template third signature, which mirrors the lower portion of the classical C-shape, would not be possible. Instead of a full reversal it is nonetheless likely that below the boundary but close to it, signs of the reversal process become observable as discussed in Sect. \ref{subsec:mac_compare}.

\subsection{Relation to macroturbulence velocities}
\label{subsec:mac_compare}

\begin{figure}
\resizebox{\hsize}{!}{\includegraphics{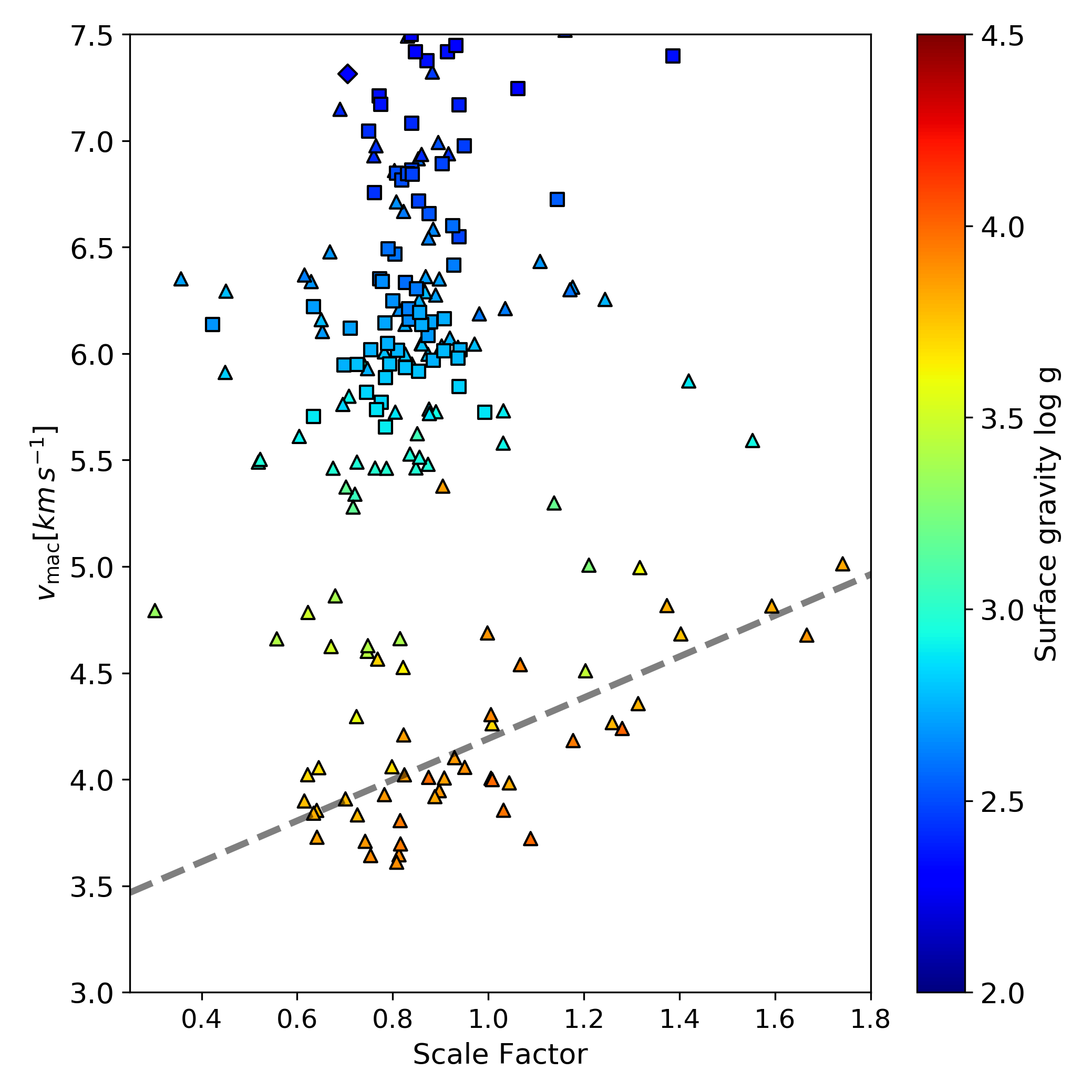}}
\caption{Comparison of macroturbulence dispersion, determined from the \citetads{2014MNRAS.444.3592D} fit, and our HARPS pMS scale factors. The gray, dashed line shows a linear fit to the subgiants with $\log g < 3.6$. Markers are colored for surface gravity (based on MIST fits). The marker shapes correspond to the evolutionary phases according to MIST: sub- or red giant (triangle), core Helium burning (squares), and asymptotic giant (diamonds).}
\label{img:v_mac}
\end{figure}

The third signature of granulation is termed as such by \citetads{2009ApJ...697.1032G} as an extension to the existing first and second signature: The macroturbulent dispersion and spectral line asymmetry respectively. They show that for their sample of eleven stars the first and third signatures are correlated, with the exception of $\iota$ Aurigae which shows an excessively high radial-tangential macroturbulence dispersion (called $v_\mathrm{mac}$ here, they called it $\zeta_\mathrm{RT}$) and a low scale factor that is well below their and our expected range at only 0.53. Since we already find a systematic difference to the \citetads{2009ApJ...697.1032G} scale factors due to differences in the derived third signature (Sec. \ref{subsec:lit_compare} and Appendix \ref{apdx:gray_investigate}) and see evidence for the same overestimation of nonempirical CBS proxies as they see for $\iota$ Aurigae (Sect. \ref{subsec:BC_comapre}), the comparison of our scale factors to macroturbulent broadening is of high interest.\par
\citetads{2014MNRAS.444.3592D} determine $v_\mathrm{mac}$ values for 28 stars from the Kepler input catalog, accounting for the influences of instrumental broadening as well as microturbulence. The effect of rotational broadening was mitigated by the use of asteroseismic $v \sin i$ values to disentangle the broadening from macroturbulence. From their results they provide a fitted relation to determine the expected $v_\mathrm{mac}$ value from effective temperature and surface gravity. Figure \ref{img:v_mac} shows the velocities obtained that way plotted against our scale factors. The picture is similar to Fig. \ref{img:BC} where the subgiant scale factors correlate well with $v_\mathrm{mac}$ and more evolved stars are expected to show a much higher velocity than the scale factors indicate. While the extent is larger, possibly because the larger number of stars in our sample highlights the deviation, this mirrors the result from $\iota$ Aurigae in \citetads{2009ApJ...697.1032G}, where $v_\mathrm{mac}$ is directly determined, rather than from a parameterized relation. It must be noted that the relation from \citetads{2014MNRAS.444.3592D} is only calibrated to MS stars, although the values obtained are within the region of those obtained by \citetads{2009ApJ...697.1032G}. The origin of this discrepancy in behavior among the more evolved stars is unclear at this point. Both works agree on the magnitudes of $v_\mathrm{mac}$ measured for MS stars as well as for the measured (Gray) and extrapolated (Doyle) values for post-MS stars. As such, and due to the consistency of the $v_\mathrm{mac}$ behavior compared to the analytically scaled CBS from Sect. \ref{subsec:BC_comapre}, we believe that the \citetads{2014MNRAS.444.3592D} relation for $v_\mathrm{mac}$ remains applicable to the giant regime. \citetads{2012AJ....143...92G} also find similar behavior for an extended sample of giant and supergiant stars where macroturbulence increased strongly with evolving luminosity class while the scale factor remains consistent for a given spectral type. A potential source discussed in both works is that the determined macroturbulence velocity depends strongly on the depth of the lines under observation, for the same reason that CBS does. While the scale factor intrinsically accounts for this through the depth-dependent third signature model, the $v_\mathrm{mac}$ values are only strictly correct for the line and star combination it is determined from because the line changes depth between stars. Even averaging multiple lines, as was done by \citetads{2014MNRAS.444.3592D}, may only partially mitigate this for small numbers of lines as we show in Appendix \ref{apdx:gray_investigate} because they can still show systematically different behavior compared to another set of lines. It is therefore possible that changes to the formation heights of the lines utilized lead to a change in $v_\mathrm{mac}$ not present in our scale factor since this shifts the sensitivity to deeper layers of the photosphere for more evolved stars. The explicit depth-dependence of the third signature avoids this effect. Alternatively, as \citetads{2012AJ....143...92G} raise for discussion, it is conceivable that for evolved stars the assumption behind the definition of macroturbulent broadening as a measure of convection (and proxy for CBS) breaks down. Macroturbulence measures a form of "convective broadening" that is dominated for the solar-like case by the granular blueshift. The changes in contrast and areal coverage mentioned in Sect. \ref{subsec:BC_comapre} could lead to an increase of the intergranular redshift contribution which would increase convection broadening from additional redshifted components while decreasing the line core shift by countering the granular velocity more strongly, effectively decoupling the first and third granulation signature for evolved stars. A more quantitative example for this can be seen in \citetads{2010ApJ...721..670G} who introduces the flux deficit as the difference between the pseudo-bisector created by the third signature and the actual bisector measurement. This flux deficit is identified as the contribution from the intergranular lanes which induces the asymmetry and its change in strength and position in velocity space, as also shown in \citetads{2010ApJ...710.1003G}, is used to explain the reversal of the C-shaped bisector of $\gamma$ Cyg. Besides the reversal of the bisector, the change in the flux deficit also serves to broaden the overall line profile which, if not explicitly taken into account, likely contributes to the increase in macroturbulent velocity toward higher temperatures and luminosities, whereas the line core and with it the third signature scaling remains largely unaffected until much closer to the granulation boundary.

\subsection{Convection velocities from 3D MHD}
\label{subsec:MHD}
\begin{figure}
\resizebox{\hsize}{!}{\includegraphics{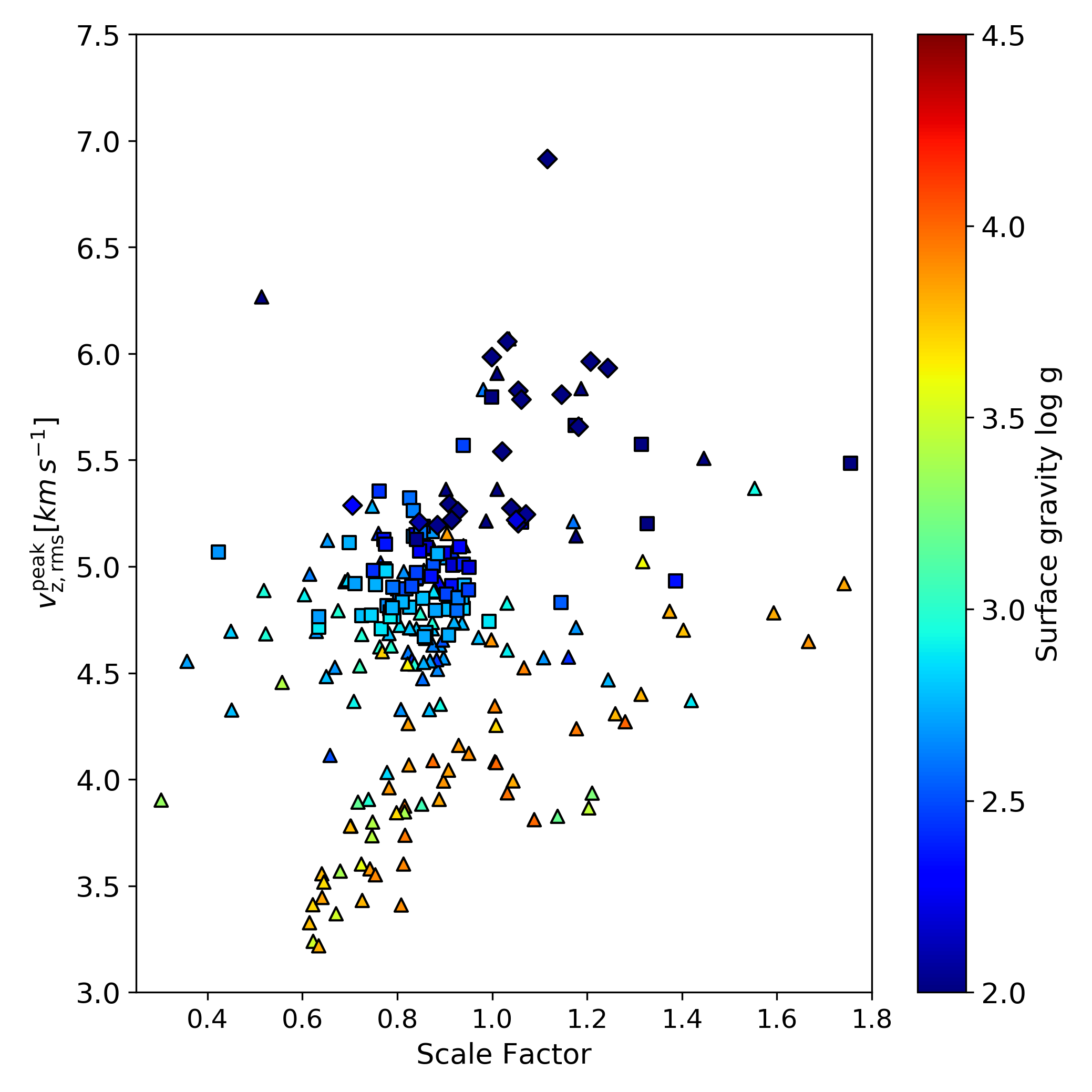}}
\caption{Comparison of horizontally averaged, peak vertical velocities from \citetads{2013A&A...557A..26M} fits to STAGGER 3D MHD model atmospheres. Markers are colored for surface gravity (based on MIST fits). The marker shapes correspond to the evolutionary phases according to MIST: sub- or red giant (triangle), core Helium burning (squares), and asymptotic giant (diamonds).}
\label{img:v_peak}
\end{figure}

We have now identified two alternatives to the scale factors determined in this work that systemically deviate in similar and potentially related ways. A third option is to investigate the actual convection velocity of the material within the convection zone. These velocities can be extracted from 3D MHD simulations such as from \citetads{2013A&A...557A..26M}. From their STAGGER grid of 3D model atmospheres they provide the horizontally averaged, peak vertical velocities $v_\mathrm{z, rms}^\mathrm{peak}$ together with a functional fit dependent on metallicity, temperature, and surface gravity. Under the assumption of solar metallicity for our stars for simplicity, since the influence is significantly weaker than for the other two parameters, we obtain Figure \ref{img:v_peak}. Once again we see the structure of more evolved stars clustered above the well correlated subgiants, although this time the tail pointing upward is missing in favor of a denser cluster. The correlation between the scale factor and $v_\mathrm{z, rms}^\mathrm{peak}$ for the subgiants also appears less strict than for the previous two comparisons with those that show weaker CBS turning slightly downward. It should be noted that $v_\mathrm{z, rms}^\mathrm{peak}$ denotes the peak velocity within the atmosphere model which extends below the photosphere and as such will overestimate the maximum possible velocity component within the line forming region. \citetads{2013A&A...557A..26M} mention this but claim that the peak velocity still scales in accordance with the global velocity magnitudes. On the other hand they state that surface gravity has an effect on the symmetry of the velocity curve which changes the velocities encountered at given optical depths because the peak is shifted to higher optical depths. This may lead to lower velocities within the line forming region, albeit the peak velocity keeps increasing. Furthermore, they use an rms value as their horizontal average so the cancellation between rising and sinking material is excluded and a bias compared to only the granular regions introduced. Similar to the fixed cancellation factor in Sect. \ref{subsec:BC_comapre} and the decoupling of convective broadening from the core shift in Sect. \ref{subsec:mac_compare}, this would lead to on overestimation of observable CBS velocities within evolved stars where the convective pattern changes drastically to a more even distribution between rising and falling material. A more definitive answer to this question could be reached with radiative transfer simulations on the atmosphere models to investigate the actual formation depths of spectral lines but this far exceeds the scope of the present work.

\section{Summary and Conclusion}
We investigated 242 post-MS stars with the third-signature scaling technique, to determine the largest cohesive sample of CBS strengths for post-MS stars to date. The technique has previously proven to be robust over a large range of temperatures for MS stars where it combines the quality advantage of an ultra-high quality solar template with the increased ease to obtain lower quality spectra. We expand on the previous results in several ways.
\begin{itemize}
\item We confirm that the technique is applicable reliably and without issues for post-MS stars and thereby extend the confirmed range of universal scaling of the third signature.
\item Our results qualitatively confirm findings by \citetads{2009ApJ...697.1032G}, that post-MS stars show increased CBS compared to MS stars of similar temperature. Quantitative differences are due to the choice of line lists and the resulting fundamentally different third-signature template, which demonstrates a different scaling behavior.
\item We determine that post-MS stars form a type of horizontal branch in scale factor versus effective temperature space. This gradually increases the CBS strength difference toward cooler temperatures, lower surface gravity, and further evolved stars, as the MS stars continue to decrease in CBS.
\item We find that subgiants gradually turn off of the MS relation toward the horizontal branch in scale factors of the post-MS stars after a slight minimum around $\log\,g \approx 3.2$.
\item This minimum may indicate a sweet spot for planet searches as it coincides with a minimum in RV jitter \citepads{2020AJ....159..235L}.
\item The near constant scale factor indicates that CBS changes very little between evolved stars which disagrees with findings from (i) analytical scaling \citep{Basu2017}, (ii) macroturbulent broadening \citepads{2014MNRAS.444.3592D}, and (iii) MHD convection velocities \citepads{2013A&A...557A..26M}. It is likely that this is due to the disconnect between line core shifts, susceptible to a fine balance between granular blue and intergranular redshift that the three listed proxies are insensitive towards.
\end{itemize}
With this paper we have demonstrated again the ease of use and generality of the third-signature scaling technique for measuring CBS. It is very robust and applicable on a wide range of not only effective temperature but also stellar evolutionary stage and, after a simple calibration for differences in instrumental broadening, the instrument used to record the data. This establishes the applicability to a very wide range of use cases which encompasses all stars with convective envelopes observed at medium resolution or higher.

\begin{acknowledgements}
We thank Onno Pols (Radboud University, Nijmegen) for useful discussions and the anonymous referee for their comments that have helped to improve the clarity of this manuscript. FL and SVJ acknowledge the support of the DFG priority program SPP 1992 "Exploring the Diversity of Extrasolar Planets (FL: RE 1664/18 and JE 701/5-1; SVJ: JE 701/5-1). The SIMBAD database\footnote{\url{http://simbad.u-strasbg.fr/simbad/}}, hosted at the CDS, Strasbourg, France, was used in this research. This research has made use of NASA's Astrophysics Data System Bibliographic Services\footnote{\url{http://adsabs.harvard.edu/}}. This work has made use of the VALD database\footnote{\url{http://vald.astro.uu.se/}}, operated at Uppsala University, the Institute of Astronomy RAS in Moscow, and the University of Vienna. This work has made use of data from the European Space Agency (ESA) mission \textit{Gaia}\footnote{\url{https://www.cosmos.esa.int/gaia}}, processed by the \textit{Gaia} Data Processing and Analysis Consortium (DPAC)\footnote{\url{https://www.cosmos.esa.int/web/gaia/dpac/consortium}}. Funding for the DPAC has been provided by national institutions, in particular the institutions participating in the \textit{Gaia} Multilateral Agreement. We thank Trifon Trifonov for providing the coadded spectra they created to be used in this work. The original observations were collected at the European Organization for Astronomical Research in the Southern Hemisphere under ESO programmes: 0100.C-0097, 0100.C-0111, 0100.C-0414,0100.C-0474, 0100.C-0487, 0100.C-0750, 0100.C-0808, 0100.C-0836, 0100.C-0847,  0100.C-0884,  0100.C-0888,  0100.D-0444,  0100.D-0717,  0101.C-0232,0101.C-0274, 0101.C-0275, 0101.C-0379, 0101.C-0407, 0101.C-0516, 0101.C-0829,  0101.D-0717,  0102.C-0338,  0102.D-0717,  0103.C-0548,  0103.D-0717,060.A-9036,  060.A-9700,  072.C-0096,  072.C-0388,  072.C-0488,  072.C-0513,072.C-0636, 072.D-0286, 072.D-0419, 072.D-0707, 073.A-0041, 073.C-0733,073.C-0784, 073.D-0038, 073.D-0136, 073.D-0527, 073.D-0578, 073.D-0590,074.C-0012,  074.C-0037,  074.C-0102,  074.C-0364,  074.D-0131,  074.D-0380,075.C-0140,  075.C-0202,  075.C-0234,  075.C-0332,  075.C-0689,  075.C-0710,075.D-0194, 075.D-0600, 075.D-0614, 075.D-0760, 075.D-0800, 076.C-0010,076.C-0073,  076.C-0155,  076.C-0279,  076.C-0429,  076.C-0878,  076.D-0103,076.D-0130, 076.D-0158, 076.D-0207, 077.C-0012, 077.C-0080, 077.C-0101,077.C-0295, 077.C-0364, 077.C-0530, 077.D-0085, 077.D-0498, 077.D-0633,077.D-0720,  078.C-0037,  078.C-0044,  078.C-0133,  078.C-0209,  078.C-0233,078.C-0403, 078.C-0751, 078.C-0833, 078.D-0067, 078.D-0071, 078.D-0245,078.D-0299,  078.D-0492,  079.C-0046,  079.C-0127,  079.C-0170,  079.C-0329,079.C-0463,  079.C-0488,  079.C-0657,  079.C-0681,  079.C-0828,  079.C-0927,079.D-0009, 079.D-0075, 079.D-0118, 079.D-0160, 079.D-0462, 079.D-0466,080.C-0032,  080.C-0071,  080.C-0664,  080.C-0712,  080.D-0047,  080.D-0086,080.D-0151, 080.D-0318, 080.D-0347, 080.D-0408, 081.C-0034, 081.C-0119,081.C-0148,  081.C-0211,  081.C-0388,  081.C-0774,  081.C-0779,  081.C-0802,081.C-0842, 081.D-0008, 081.D-0065, 081.D-0109, 081.D-0531, 081.D-0610,081.D-0870,  082.B-0610,  082.C-0040,  082.C-0212,  082.C-0308,  082.C-0312,082.C-0315,  082.C-0333,  082.C-0357,  082.C-0390,  082.C-0412,  082.C-0427,082.C-0608,  082.C-0718,  083.C-0186,  083.C-0413,  083.C-0794,  083.C-1001,  083.D-0668,  084.C-0185,  084.C-0228,  084.C-0229,  084.C-1039,  085.C-0019,085.C-0063,  085.C-0318,  085.C-0393,  086.C-0145,  086.C-0230,  086.C-0284,086.D-0240,  087.C-0012,  087.C-0368,  087.C-0649,  087.C-0831,  087.C-0990,087.D-0511,  088.C-0011,  088.C-0323,  088.C-0353,  088.C-0513,  088.C-0662,089.C-0006,  089.C-0050,  089.C-0151,  089.C-0415,  089.C-0497,  089.C-0732,089.C-0739,  090.C-0395,  090.C-0421,  090.C-0540,  090.C-0849,  091.C-0034,091.C-0184,  091.C-0271,  091.C-0438,  091.C-0456,  091.C-0471,  091.C-0844,091.C-0853,  091.C-0866,  091.C-0936,  091.D-0469,  092.C-0282,  092.C-0454,092.C-0579,  092.C-0721,  092.C-0832,  092.D-0261,  093.C-0062,  093.C-0409,093.C-0417,  093.C-0474,  093.C-0919,  094.C-0090,  094.C-0297,  094.C-0428,094.C-0797,  094.C-0894,  094.C-0901,  094.C-0946,  094.D-0056,  094.D-0596,095.C-0040,  095.C-0105,  095.C-0367,  095.C-0551,  095.C-0718,  095.C-0799,095.C-0947,  095.D-0026,  095.D-0717,  096.C-0053,  096.C-0082,  096.C-0183,096.C-0210,  096.C-0331,  096.C-0417,  096.C-0460,  096.C-0499,  096.C-0657,096.C-0708,  096.C-0762,  096.C-0876,  096.D-0402,  096.D-0717,  097.C-0021,097.C-0090,  097.C-0390,  097.C-0434,  097.C-0561,  097.C-0571,  097.C-0864,097.C-0948,  097.C-1025,  097.D-0156,  097.D-0717,  098.C-0269,  098.C-0292,098.C-0304,  098.C-0366,  098-C-0518,  098.C-0518,  098.C-0739,  098.C-0820,098.C-0860,  098.D-0717,  099.C-0093,  099.C-0138,  099.C-0205,  099.C-0303,099.C-0304,  099.C-0374,  099.C-0458,  099.C-0491,  099.C-0798,  099.C-0880,099.C-0898, 099.D-0717, 1101.C-0721, 180.C-0886, 183.C-0437, 183.C-0972,183.D-0729,  184.C-0639,  184.C-0815,  185.D-0056,  188.C-0265,  188.C-0779,190.C-0027,  191.C-0505,  191.C-0873,  192.C-0224,  192.C-0852,  196.C-0042,196.C-1006,  198.C-0169,  198.C-0836,  198.C-0838,  281.D-5052,  281.D-5053,282.C-5034,  282.C-5036,  282.D-5006,  283.C-5017,  283.C-5022,  288.C-5010,292.C-5004, 295.C-5031, 495.L-0963, 60.A-9036, 60.A-9700, and 63.A-9036.
The analysis was carried out using the programming language Python3\footnote{\url{https://www.python.org/}} Version 3.7.6 \citep{10.5555/1593511}, and the accompanying software packages: Numpy\footnote{\url{https://numpy.org/}} Version 1.18.1 \citep{harris2020array}, Scipy\footnote{\url{https://www.scipy.org/scipylib/}} Version 1.4.1 \citep{2020SciPy-NMeth}, Astropy\footnote{\url{https://www.astropy.org/}} Version 4.0 \citep{astropy:2013, astropy:2018}, Astroquery\footnote{\url{https://astroquery.readthedocs.io}} Version 0.4.3 \citep{2019AJ....157...98G}, and Matplotlib\footnote{\url{https://matplotlib.org/}} Version 3.1.3 \citep{Hunter:2007}.
\end{acknowledgements}

\bibliographystyle{aa} 
\bibliography{paper} 

\begin{appendix}

\section{Further investigations}
\label{apdx:gray_investigate}
\begin{figure}
\resizebox{\hsize}{!}{\includegraphics{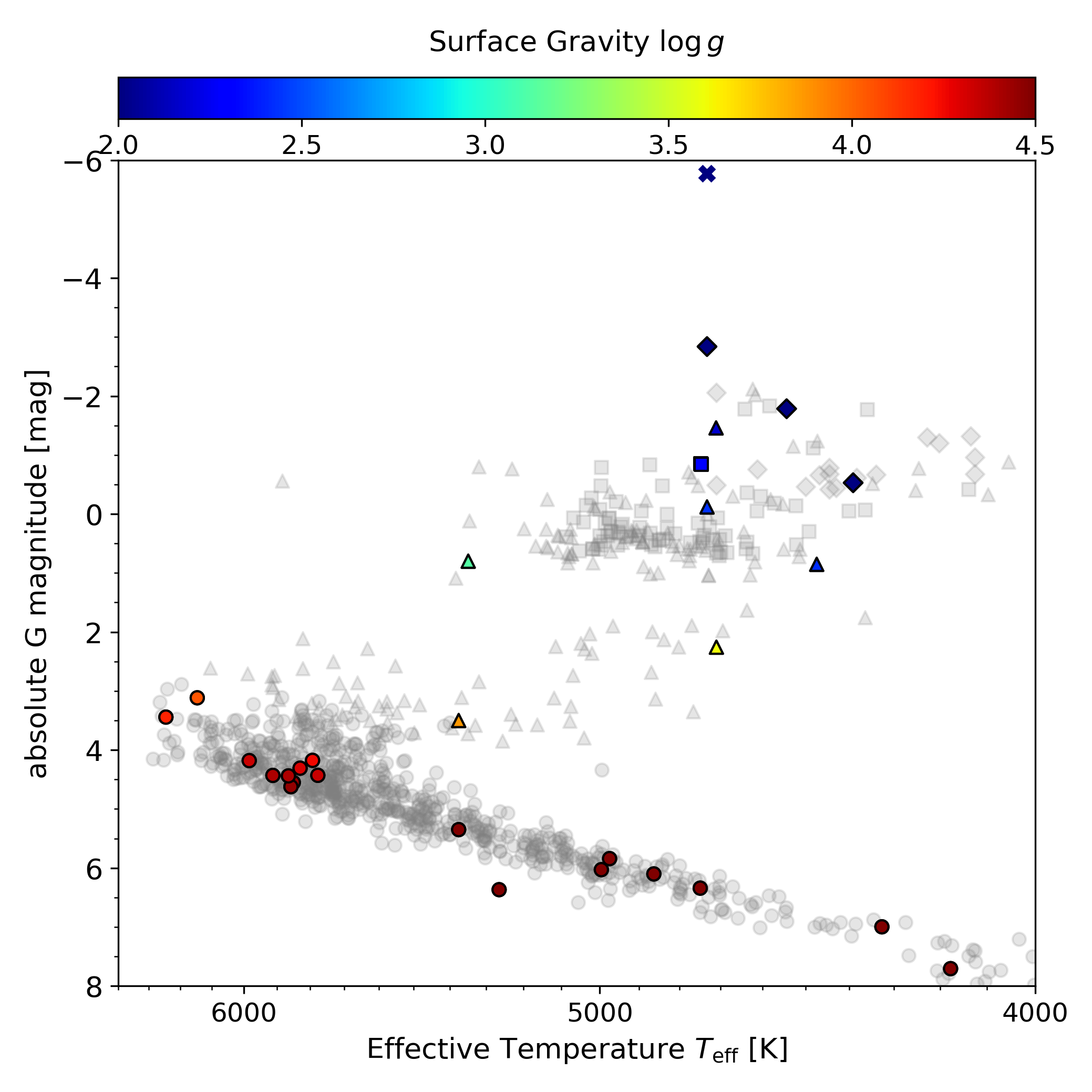}}
\caption{Hertzsprung-Russel diagram for our control sample of PEPSI stars based on \textit{Gaia} DR2 data and colored for surface gravity (based on MIST fits). The HARPS MS/post-MS sample from paper I and this work is included as reference (gray markers). Three PEPSI stars are not shown due to missing \textit{Gaia} parallaxes. The marker shapes correspond to the evolutionary phases according to MIST: main sequence (circle), sub- or red giant (triangle), core Helium burning (squares), asymptotic giant (diamonds), extended asymptotic giant (stars), and thermally pulsing asymptotic giant (crosses).}
\label{img:PEPSI_HR}
\end{figure}

\begin{figure}
\resizebox{\hsize}{!}{\includegraphics{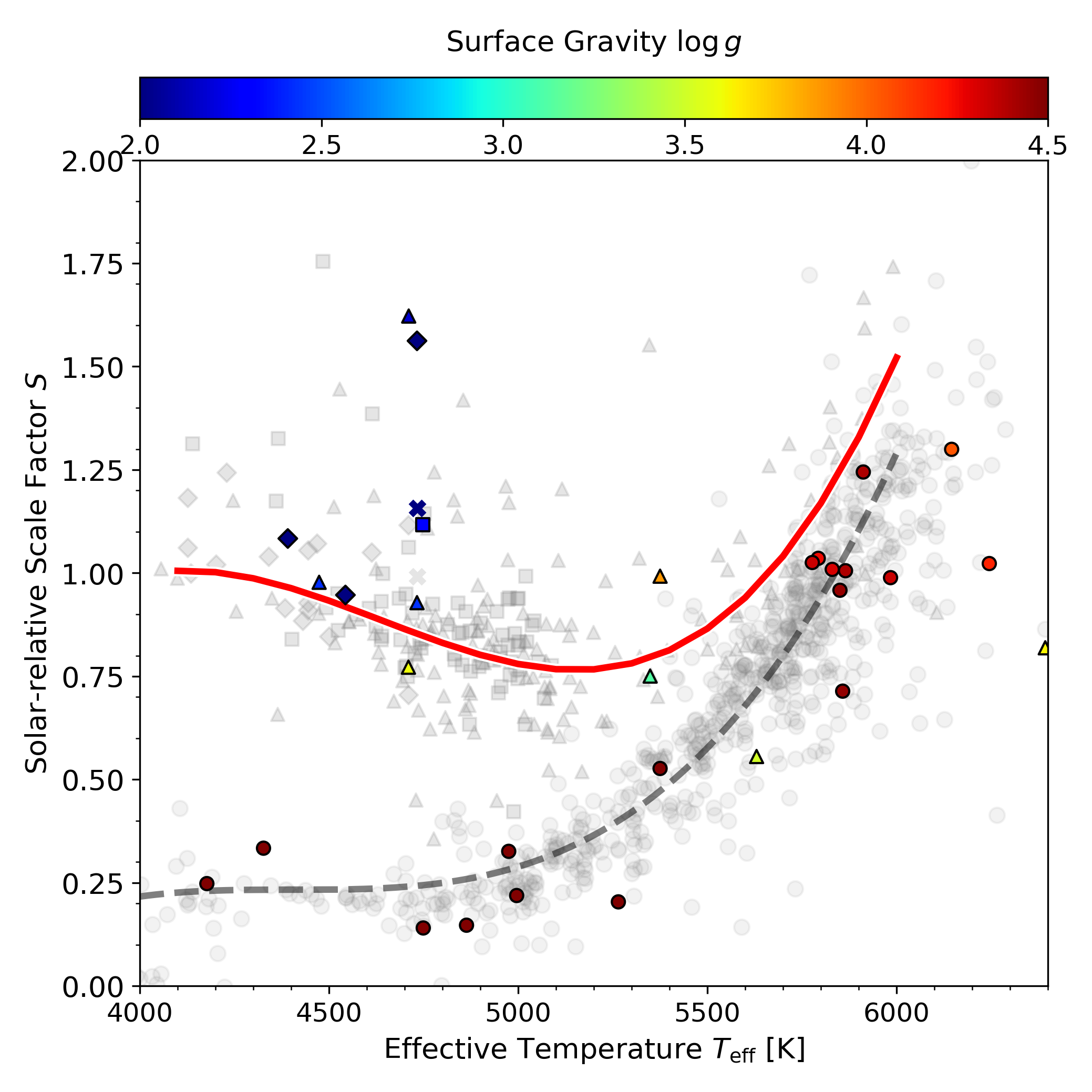}}
\caption{Scale factor results for the PEPSI control sample (colored for MIST surface gravity). Both MS and post-MS scales are in agreement with the main HARPS sample models (gray and red curves for the fitted models, gray markers for individual stars from the samples of paper I and this work). The marker shapes correspond to the evolutionary phases according to MIST: main sequence (circle), sub- or red giant (triangle), core Helium burning (squares), asymptotic giant (diamonds), extended asymptotic giant (stars), and thermally pulsing asymptotic giant (crosses).}
\label{img:PEPSI_scales}
\end{figure}

We find that quantitatively our results for the scale factor and those from \citetads{2009ApJ...697.1032G} differ by up to a factor of three (S$\sim$0.27 versus S$\sim$0.72 for the late K dwarf $\epsilon$ Eridani). In this section we seek to find the source of this discrepancy. First, we looked for stars included in his sample as well as ours for a direct comparison. This leaves us with $\epsilon$ Eridani and $\tau$ Ceti, both MS stars that show significantly lower scale factors in our analysis, though $\tau$ Ceti deviates only by a factor of two (0.56 versus 1.01). We endeavored to exclude an instrumental source for the difference, albeit that \citetads{2009ApJ...697.1032G} cites a resolving power of his spectra comparable to HARPS, and analyzed a number of published PEPSI spectra from \citetads{2018A&A...612A..45S} (HR diagram for the sample in Fig. \ref{img:PEPSI_HR}). The results, as shown in Fig. \ref{img:PEPSI_scales}, agree with our HARPS sample within the margin of error, both for MS and post-MS stars, and reinforce the systematic difference to \citetads{2009ApJ...697.1032G}.\par
\begin{figure}
\resizebox{\hsize}{!}{\includegraphics{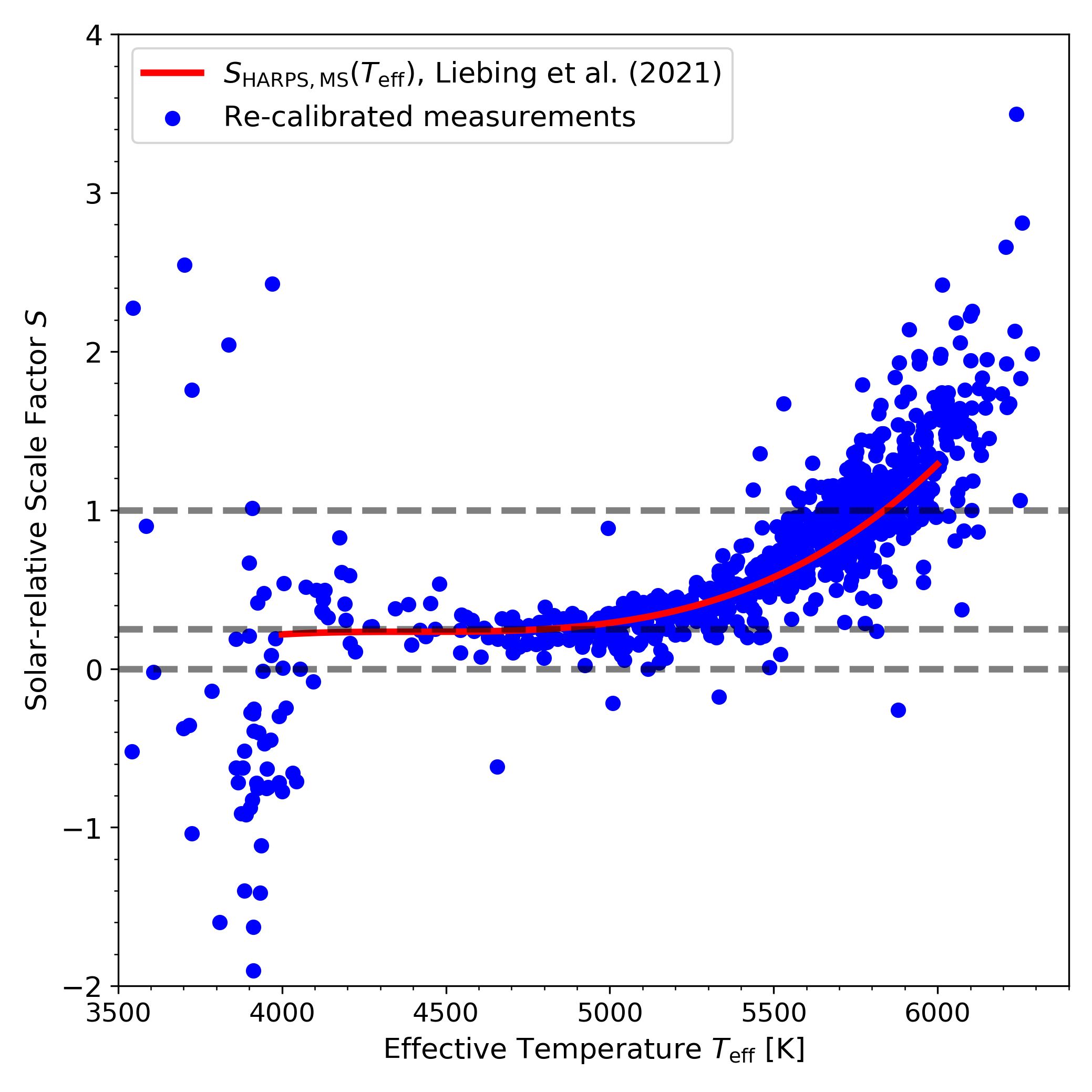}}
\caption{Scale factor vs. effective temperature for the HARPS MS sample from paper I, using the 15 lines from \citetads{2009ApJ...697.1032G} and our solar 3rd signature, prescaled by a factor of 2.1 to account for the systematic difference due to the line choice. The resulting recalibrated scale factors (blue dots) are indistinguishable from the fit to the results from paper I (red line, Eq. (3) from paper I). The dashed lines mark scale factors of zero, one (solar strength), and 0.25 (K dwarf plateau from paper I).}
\label{img:HARPS_GrayLines}
\end{figure}

Besides the instrument, which we have shown to not be the source, another difference lies in the choice of spectral lines. Where we use a general list, not specifically curated to any degree and only vetted to optimize third-signature fit quality among the entire sample over all spectral types, with a total of over 1000 lines, \citetads{2009ApJ...697.1032G} uses a list of 15 lines of \ion{Fe}{I} that are easy to measure in the Sun. As we point out in paper I, from our own experience, care must be taken to ensure a list of lines is generally applicable and not overly adapted to any specific type of star, lest the results become unreliable once applied to an unsuited star. We also show for the sun that the wavelength range covered by an otherwise representative line list influences the shape and strength of the third signature which affects the scaling behavior. To check if either was the case here, we repeated our analysis with the line list provided by \citetads{2009ApJ...697.1032G}, but retained our third-signature model (Eq. \ref{eqn:thirdSigModel}). The results are shown in Fig. \ref{img:HARPS_GrayLines}. Besides a generally larger scale factor that could be calibrated for (solar fit at S$\sim$2.1), they agree again with our results from paper I. While the uncertainties are larger, as is the scatter of individual stars, this eliminates the line list itself as the source for the deviation as well, especially as the \citetads{2009ApJ...697.1032G} MS stars also match our MS relation after recalibration. This further matches our experience with the results from \citetads{2017A&A...597A..52M} and \citetads{2017A&A...607A.124M}, where general agreement is reached with the results in paper I, despite a significant difference in line list. Overall this points to the second previously mentioned point that the line choice is representative for CBS strength but differs in its scaling behavior.\par

\begin{figure}
\resizebox{\hsize}{!}{\includegraphics{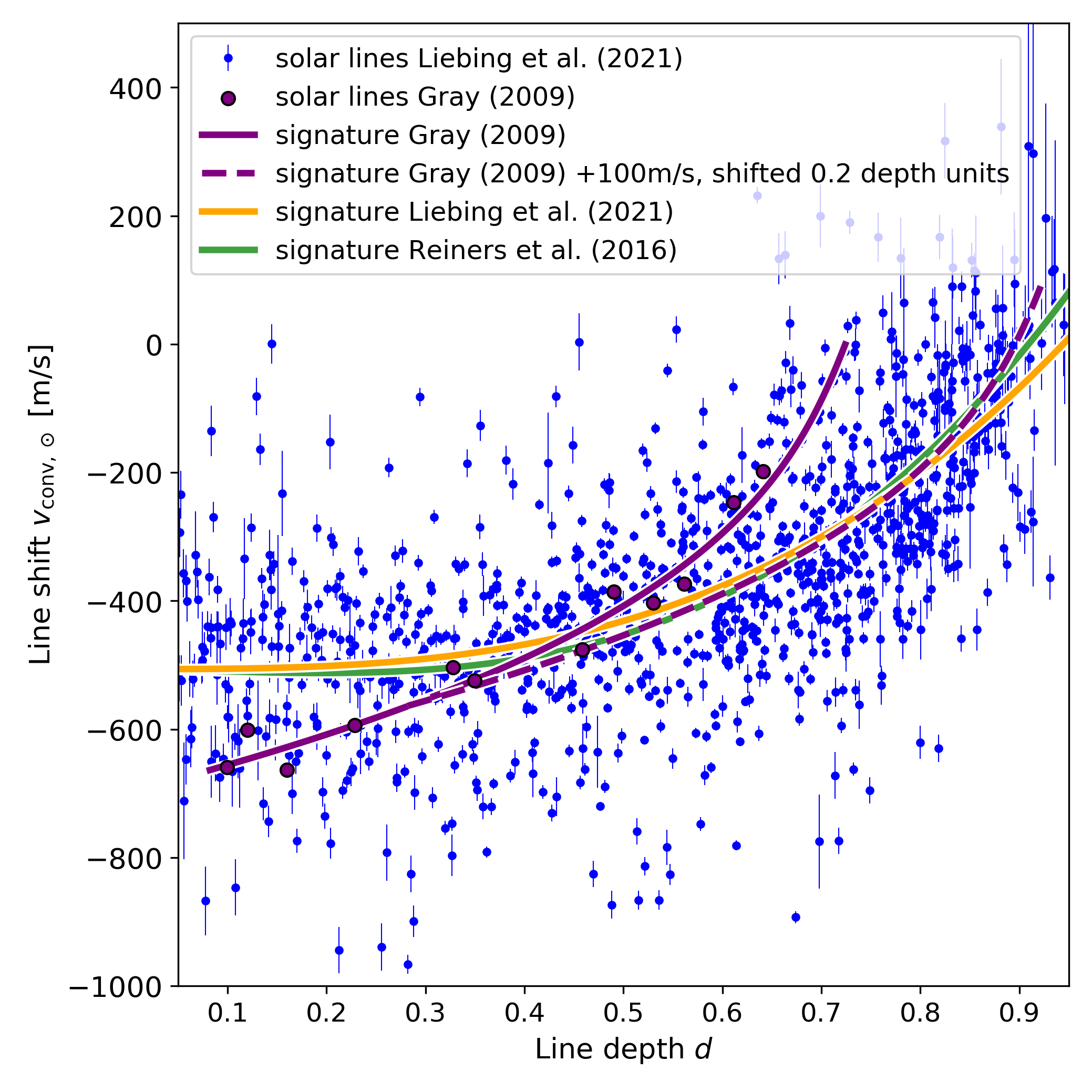}}
\caption{Comparison of the \citetads{2009ApJ...697.1032G} line measurements and standard curve $F/F_\mathrm{c}\left(v\right)$ for the Sun with our equivalent third signature $v_\mathrm{conv,\odot}\left(d\right)$ determined in paper I (Eq. \ref{eqn:thirdSigModel}, see also Fig. \ref{img:3rdsigexmpl} of this work) and the \citetads{2016A&A...587A..65R} signature.}
\label{img:Gray_SC_comparison}
\end{figure}

To check this, we compared our solar-based third-signature model to the one used by \citetads{2009ApJ...697.1032G} and \citetads{2018ApJ...852...42G}. We applied a correction to our model in the form of an RV shift to account for the different treatment of barycentric motion and solar gravitational redshift and compared the two models. However, the difference in absolute RV shift is not the source of the differences in the results as an absolute shift is already a fitted parameter for the third signature scaling and, unlike the CBS strength itself, not a differential shift. As such it does not change the determined CBS strength and can not account for the difference in results. A direct comparison of the two models in Fig. \ref{img:Gray_SC_comparison} reveals a clear difference in that the \citetads{2009ApJ...697.1032G} model curves up, toward weaker CBS, for significantly shallower lines. If one shifts that model by 20 percentage points toward deeper lines and adjusts the absolute RV, this results in an almost perfect match to our model and the one from \citetads{2016A&A...587A..65R}. We speculate however, that this offset is not due to actual differences in the spectra or measurements of the line depths, but rather a result of the choice of lines as mentioned previously which lead to a different but equally valid third-signature template. A direct comparison of the measurements from our technique but using the line list from \citetads{2009ApJ...697.1032G} against the extracted line data from Fig. 7 from \citetads{2009ApJ...697.1032G} for the Sun (a R$\sim$100\,000 degraded IAG atlas in our case) reveals a good agreement among the two and to the standard curve from their work. The fact that our template model again requires a scaling of 2.1, like the stars from the MS sample, rather than 1.0 as one would expect for a fit to solar lines, once more indicates that the third signature template (or standard curve) simply scales differently but otherwise represents the same underlying effect. This is shown in Fig. \ref{img:Gray_SC_comparison}, where we compare \citetads{2009ApJ...697.1032G} line measurements and standard curve with our equivalent line-by-line results from paper I. We clearly see the discrepancy in standard curve / third signature template is purely due to the choice of spectral lines, as all the lines chosen by \citetads{2009ApJ...697.1032G} are within the range covered by our own choice of line list. As such, the results are not in actual disagreement, despite the difference in similar sounding values, but are simply not quantitatively comparable when one does not account for the difference in scaling (as shown in Fig. \ref{img:HARPS_GrayLines}).\par
This conclusion is supported by results from \citetads{2012AJ....143..112G} and \citetads{2014ApJ...796...88G} for $\zeta$ Cygni and $\beta$ Gem respectively. While they use the same standard curve as in \citetads{2009ApJ...697.1032G}, their list of \ion{Fe}{I} lines was supplemented with additional, non-\ion{Fe}{I} lines and results in lower than expected scale factors of 1.08 and 0.8 respectively, comparable to the range of our results.

We show in paper I that there is a degeneracy between observed line depth, influenced by line broadening effects, and the obtained scale factor. This results in the necessity to calibrate the template to the resolving power of the instrument, to account for instrumental broadening, that would otherwise lead to a similar offset in measured solar CBS scale factor. This further explains the growing discrepancy in published scale factors between this work and \citetads{2009ApJ...697.1032G} for decreasing effective temperatures for $\tau$ Ceti and $\epsilon$ Eridani as the underlying signature is fundamentally different and therefore scales differently as well, in this instance much slower. As such our results and those of \citetads{2009ApJ...697.1032G} appear to both be valid within their context and simply not directly comparable. This again matches our observations from paper I against \citetads{2017A&A...597A..52M} and \citetads{2017A&A...607A.124M}, where a linear third signature model leads to similar issues, though at higher scale factors, as there the model underestimates the third signature curvature.

\section{Additional figures}
\label{apdx:add_figures}
\begin{figure}
\resizebox{\hsize}{!}{\includegraphics{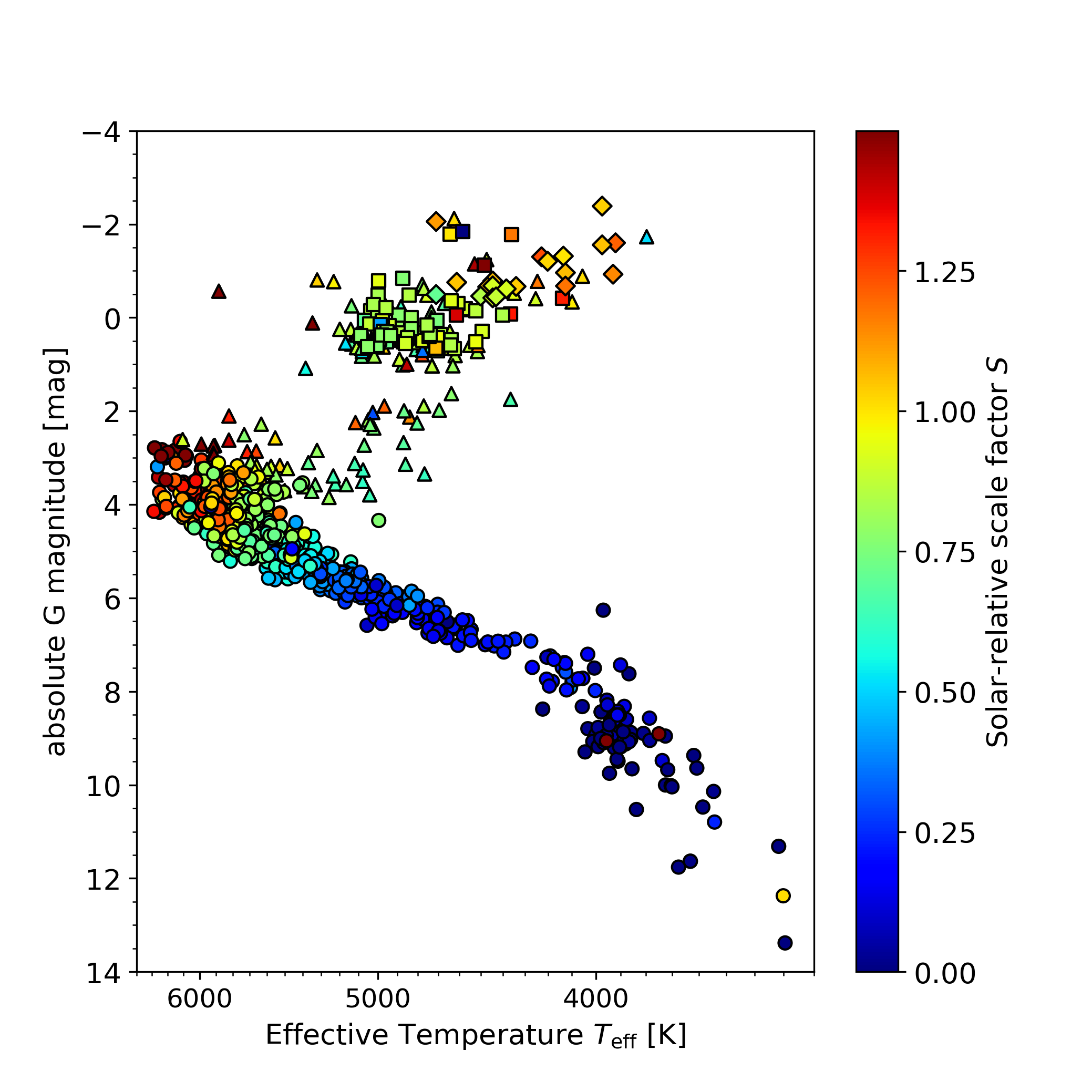}}
\caption{Combined sample from paper I and this work shown as an Hertzsprung-Russel diagram, same as Fig. \ref{img:HR} but colored for the CBS strength.}
\label{img:CBS_HR}
\end{figure}

Figure \ref{img:CBS_HR} shows the changes in observed CBS strength over the combined samples from paper I and this work, similar to Figure \ref{img:scalesplot} but in the form of an Hertzsprung-Russel diagram (See Fig. \ref{img:HR}).

\section{Tables}
For ease of access and a quicker overview, we list in Table \ref{tab:cheatsheet} the expected CBS strengths and velocities for the range of 4100\,K - 6000\,K based on our model from Eq. \ref{eqn:SG_fit}. Table \ref{tab:star_data} lists all stars of our sample with basic parameters important to this work and the determined CBS strengths and uncertainties. The full Table is available via CDS.
\begin{table}
\centering
\caption{Scale factor cheat sheet following Eq. \ref{eqn:SG_fit}. Available online at CDS.\label{tab:cheatsheet}}
\begin{tabular}{c c c c}
\hline\hline
Effective temperature [K] & Scale factor & Velocity [m\,s$^{-1}$]\\
\hline
4100 & 1.005 & -361.3\\
4200 & 1.002 & -360.1\\
4300 & 0.987 & -354.7\\
4400 & 0.963 & -346.2\\
4500 & 0.933 & -335.3\\
4600 & 0.899 & -323.1\\
4700 & 0.864 & -310.6\\
4800 & 0.831 & -298.6\\
4900 & 0.802 & -288.1\\
5000 & 0.780 & -280.2\\
5100 & 0.767 & -275.6\\
5200 & 0.767 & -275.5\\
5300 & 0.781 & -280.7\\
5400 & 0.813 & -292.2\\
5500 & 0.865 & -311.0\\
5600 & 0.940 & -337.9\\
5700 & 1.041 & -374.0\\
5800 & 1.169 & -420.2\\
5900 & 1.329 & -477.5\\
6000 & 1.521 & -546.8\\
\hline
\end{tabular}
\tablefoot{The velocities given assume a line depth of 0.7 (Eq. \ref{eqn:thirdSigModel}), corresponding to a median solar line.}
\end{table}

\longtab[2]{
\begin{landscape}
\begin{longtable}{cccccccccccccc}
\caption{List of all stars with their parameters.\label{tab:star_data}}\\
\hline\hline
Star & DR2 ID & $T_\mathrm{eff}$ & $G_\mathrm{mag}$ & $\log g$ & Phase & $v\cdot\sin i$ & \#Spectra & S/N & \#Lines & $\chi^2_\mathrm{P}$ & Scale $S$ & $\sigma_S$\\
\hline
\endfirsthead
\caption{continued.}\\
\hline\hline
Star & DR2 ID & $T_\mathrm{eff}$ & $G_\mathrm{mag}$ & $\log g$ & Phase & $v\cdot\sin i$ & \#Spectra & S/N & \#Lines & $\chi^2_\mathrm{P}$ & Scale $S$ & $\sigma_S$\\
\hline
\endhead
\hline
\endfoot
HD56618 & 5606936597846400512 & $3797$ & $-1.73$ & $0.90$ & RGB & $-$ & $5$ & $544$ & $923$ & $0.92$ & $0.51$ & $0.12$\\
HD41047 & 2889305249701803264 & $3920$ & $-1.60$ & $1.15$ & AGB & $-$ & $9$ & $462$ & $1095$ & $0.03$ & $1.21$ & $0.07$\\
HD12524 & 4944212458182232192 & $3930$ & $-0.93$ & $1.22$ & AGB & $-$ & $9$ & $523$ & $1118$ & $0.24$ & $1.15$ & $0.06$\\
HD61772 & 3029166186779364736 & $3974$ & $-2.39$ & $1.17$ & AGB & $-$ & $35$ & $1347$ & $1141$ & $0.07$ & $1.03$ & $0.07$\\
HD45669 & 5278749385021344640 & $3974$ & $-1.56$ & $1.26$ & AGB & $-$ & $7$ & $353$ & $1109$ & $0.03$ & $1.06$ & $0.06$\\
HD130694 & 6219380139468557568 & $4055$ & $-0.89$ & $1.54$ & RGB & $-$ & $194$ & $3039$ & $1165$ & $0.03$ & $1.01$ & $0.04$\\
BD-184705 & 4119382783830154112 & $4098$ & $-0.33$ & $1.66$ & RGB & $-$ & $13$ & $160$ & $1053$ & $0.27$ & $0.99$ & $0.08$\\
TYC5126-2775-1 & 4252502954347971840 & $4126$ & $-0.68$ & $1.50$ & AGB & $4.50$ & $8$ & $88$ & $1013$ & $0.05$ & $1.18$ & $0.13$\\
HD50778 & 2952868119981800448 & $4126$ & $-0.96$ & $1.44$ & AGB & $3.60$ & $7$ & $392$ & $1122$ & $0.01$ & $1.06$ & $0.05$\\
HD28028 & 4871775051596978944 & $4135$ & $-1.32$ & $1.37$ & AGB & $-$ & $7$ & $474$ & $1121$ & $0.04$ & $1.00$ & $0.07$\\
TYC8963-1543-1 & 5336820469612710784 & $4139$ & $-0.42$ & $1.54$ & CHeB & $-$ & $21$ & $184$ & $1084$ & $0.48$ & $1.31$ & $0.05$\\
HD170053 & 4477460391998886144 & $4202$ & $-1.21$ & $1.63$ & AGB & $-$ & $102$ & $1433$ & $1172$ & $0.04$ & $1.02$ & $0.06$\\
BD+101802 & 600784874384342400 & $4229$ & $-1.30$ & $1.49$ & AGB & $-$ & $3$ & $247$ & $1034$ & $0.53$ & $1.24$ & $0.10$\\
HD45398 & 3104552174309524096 & $4246$ & $-0.78$ & $1.85$ & RGB & $-$ & $105$ & $1648$ & $1169$ & $0.02$ & $1.18$ & $0.06$\\
BD-184710 & 4119382852549575168 & $4254$ & $-0.40$ & $1.90$ & RGB & $-$ & $16$ & $220$ & $1091$ & $0.11$ & $0.91$ & $0.08$\\
HD122430 & 6177432396478424576 & $4340$ & $-0.67$ & $1.89$ & AGB & $2.59$ & $7$ & $370$ & $1157$ & $0.03$ & $1.04$ & $0.05$\\
CD-546003 & 5887641435423836544 & $4348$ & $-0.52$ & $1.98$ & RGB & $-$ & $7$ & $185$ & $1104$ & $0.04$ & $0.94$ & $0.06$\\
HD96789 & 5340186143451130112 & $4360$ & $-1.77$ & $1.74$ & CHeB & $-$ & $7$ & $339$ & $1122$ & $0.31$ & $1.17$ & $0.07$\\
COROT102387834 & 3317885502001094400 & $4364$ & $1.75$ & $2.50$ & RGB & $-$ & $17$ & $64$ & $1007$ & $5.29$ & $0.66$ & $0.12$\\
HD10615 & 4715993838992279680 & $4365$ & $-0.07$ & $1.95$ & CHeB & $-$ & $9$ & $366$ & $1158$ & $0.02$ & $1.33$ & $0.03$\\
HD47536 & 5583831735369515008 & $4384$ & $-0.62$ & $1.96$ & AGB & $1.81$ & $11$ & $485$ & $1182$ & $0.03$ & $0.91$ & $0.04$\\
HD85859 & 5659110864115634944 & $4401$ & $-0.06$ & $2.02$ & CHeB & $-$ & $10$ & $703$ & $1145$ & $0.04$ & $0.84$ & $0.04$\\
HD65695 & 3068999225306100480 & $4430$ & $-0.45$ & $2.02$ & AGB & $2.00$ & $10$ & $568$ & $1189$ & $0.07$ & $0.88$ & $0.05$\\
HD110014 & 3676091134604409728 & $4445$ & $-0.78$ & $2.04$ & AGB & $3.90$ & $111$ & $2253$ & $1187$ & $0.01$ & $1.05$ & $0.05$\\
HD178484 & 4261348186123153408 & $4445$ & $-0.42$ & $2.01$ & AGB & $-$ & $146$ & $1880$ & $1201$ & $0.03$ & $0.93$ & $0.05$\\
HD1690 & 2430036837596487424 & $4445$ & $-0.68$ & $2.00$ & AGB & $3.02$ & $44$ & $363$ & $1141$ & $0.07$ & $0.91$ & $0.05$\\
HD132944 & 5887670572485812608 & $4468$ & $-0.66$ & $2.04$ & AGB & $-$ & $8$ & $239$ & $1101$ & $0.08$ & $1.07$ & $0.06$\\
HD206642 & 6585580939594718464 & $4472$ & $-1.24$ & $1.99$ & RGB & $-$ & $3$ & $525$ & $1163$ & $0.06$ & $0.90$ & $0.04$\\
HD78004 & 5330160246630987776 & $4483$ & $-1.13$ & $1.95$ & CHeB & $-$ & $4$ & $420$ & $1173$ & $0.87$ & $1.75$ & $0.03$\\
CD-4911415 & 5949577646955571712 & $4492$ & $0.30$ & $2.18$ & CHeB & $-$ & $34$ & $234$ & $1137$ & $0.06$ & $0.92$ & $0.04$\\
HD171759 & 6419476755915457792 & $4500$ & $-0.46$ & $2.09$ & AGB & $-$ & $7$ & $542$ & $1180$ & $0.04$ & $0.85$ & $0.04$\\
HD23319 & 4856903813234485248 & $4512$ & $0.60$ & $2.41$ & RGB & $1.70$ & $10$ & $482$ & $1184$ & $0.02$ & $1.16$ & $0.04$\\
HD169370 & 4160006233744631040 & $4515$ & $0.73$ & $2.42$ & RGB & $4.40$ & $5$ & $219$ & $1172$ & $0.08$ & $0.83$ & $0.04$\\
HD159194 & 5961811431807564032 & $4522$ & $0.51$ & $2.21$ & CHeB & $-$ & $9$ & $232$ & $1140$ & $0.02$ & $0.95$ & $0.06$\\
CD-535815 & 5887699640826811520 & $4523$ & $-0.14$ & $2.17$ & CHeB & $-$ & $8$ & $176$ & $1115$ & $0.18$ & $0.86$ & $0.05$\\
HD1522 & 2429978803998328960 & $4528$ & $-1.15$ & $1.99$ & RGB & $0.40$ & $5$ & $491$ & $1188$ & $0.30$ & $1.45$ & $0.04$\\
HD170031 & 4476782577440717824 & $4551$ & $0.59$ & $2.46$ & RGB & $-$ & $111$ & $1362$ & $1204$ & $0.03$ & $0.88$ & $0.04$\\
HD188114 & 6687035763708302848 & $4553$ & $-0.17$ & $2.30$ & RGB & $-$ & $8$ & $578$ & $1190$ & $0.04$ & $0.88$ & $0.04$\\
BD-132130 & 3030262468592291072 & $4572$ & $-0.19$ & $2.24$ & CHeB & $2.14$ & $37$ & $234$ & $1142$ & $0.06$ & $0.90$ & $0.05$\\
CD-535820 & 5887695517658166912 & $4581$ & $-0.25$ & $2.31$ & RGB & $-$ & $8$ & $181$ & $1111$ & $0.17$ & $0.89$ & $0.06$\\
HD87833 & 5256491799378573824 & $4583$ & $-1.84$ & $1.96$ & CHeB & $-$ & $18$ & $328$ & $959$ & $0.71$ & $-0.14$ & $0.24$\\
HD200763 & 6781170726597994240 & $4605$ & $-0.31$ & $2.34$ & CHeB & $-$ & $9$ & $449$ & $1197$ & $0.05$ & $0.92$ & $0.05$\\
HD18293 & 4640087816818486528 & $4612$ & $-0.76$ & $2.21$ & AGB & $-$ & $10$ & $473$ & $1150$ & $0.04$ & $1.05$ & $0.06$\\
... & ... & ... & ... & ... & ... & ... & ... & ... & ... & ... & ... & ...\\
\end{longtable}
\end{landscape}
}

\end{appendix}
\end{document}